\documentclass{article}
\pdfoutput=1

\usepackage{arxiv}
\usepackage[utf8]{inputenc} 
\usepackage[T1]{fontenc} 
\usepackage{hyperref} 
\hypersetup{
colorlinks = true,
linkcolor = black,
anchorcolor = black,
citecolor = black,
filecolor = black,
urlcolor = blue
}
\usepackage{booktabs} 
\usepackage{amsfonts} 
\usepackage{nicefrac} 
\usepackage{microtype} 
\usepackage{amsmath,bm}
\usepackage{graphicx}
\usepackage[flushleft]{threeparttable}
\usepackage{multirow}
\usepackage{natbib}
\usepackage{makecell}
\usepackage{caption}
\usepackage{subcaption} 
\usepackage{float}

\newcommand{\figurehere}[1]{\begin{center}%
=========================\\%
Insert Figure #1 about here\\%
=========================\\%
\end{center}}
\newcommand{\tablehere}[1]{\begin{center}%
=========================\\%
Insert Table #1 about here\\%
=========================\\%
\end{center}}

\usepackage{array}
\newcommand{\PreserveBackslash}[1]{\let\temp=\\#1\let\\=\temp}
\newcolumntype{C}[1]{>{\PreserveBackslash\centering}p{#1}}
\newcolumntype{R}[1]{>{\PreserveBackslash\raggedleft}p{#1}}
\newcolumntype{L}[1]{>{\PreserveBackslash\raggedright}p{#1}}

\title{Assessing Mediational Processes Using Piecewise Linear Growth Curve Models with Individual Measurement Occasions}

\author{
Jin Liu \thanks{CONTACT Jin Liu Email: Veronica.Liu0206@gmail.com, \textcircled{c}2022, Behavior Research Methods. This paper is not the copy of record and may not exactly replicate the final, authoritative version of the article. Please do not copy or cite without authors' permission. The final article will be available, upon publication, via its DOI: \url{10.3758/s13428-022-01940-2}}\\
Department of Biostatistics\\
Virginia Commonwealth University \\
 \And
Robert A. Perera\\
Department of Biostatistics\\
Virginia Commonwealth University \\
}

\begin{document}
\maketitle
\begin{abstract}
Longitudinal processes often unfold concurrently where the growth of two or more longitudinal outcomes are associated. Additionally, if the study under investigation is long, the growth curves may exhibit nonconstant change with respect to time. Multiple existing studies have developed multivariate growth models with nonlinear functional forms to explore joint development where two longitudinal records are correlated over time. However, the relationship between multiple longitudinal outcomes may also be unidirectional. Accordingly, it is of interest to estimate regression coefficients of such unidirectional paths. One statistical tool for such analyses is longitudinal mediation models. In this study, we develop two models to evaluate mediational processes where the linear-linear piecewise growth model is utilized to capture the change patterns. We define the mediational process as either the baseline covariate or the change in covariate influencing the change in the mediator, which, in turn, affects the change in the outcome. We present the proposed models through simulation studies and real-world data analyses. Our simulation studies demonstrate that the proposed mediational models can provide unbiased and accurate point estimates with target coverage probabilities with a $95\%$ confidence interval. The empirical analyses demonstrate that the proposed model can estimate covariates' direct and indirect effects on the change in the outcome. We also provide the corresponding code for the proposed models.
\end{abstract}


\setcounter{secnumdepth}{3}
\section*{Introduction}\label{Intro}
In the studies with a longitudinal design, multiple variables of interest are collected together. For example, the primary endpoint and multiple secondary endpoints are often recorded repeatedly in clinical trials. In observational studies, such as The Early Childhood Longitudinal Study, Kindergarten Cohort: $2010-11$ (ECLS-K: $2011$) \citep{Tourangeau2013ECLS}, test scores of multiple disciplinary subjects are collected each school year. Given that a longitudinal process exhibits a nonlinear relationship with respect to time $t$ if there are some periods where change is more rapid than in others, earlier studies, such as \citet{Blozis2004MGM, Blozis2008MGM, Peralta2020PBLSGM, Liu2021PBLSGM}, have developed multivariate growth models (MGMs) \citep[Chapter~8]{Grimm2016growth} to explore joint nonlinear developmental processes, with a focus on an association between these processes by estimating the covariances of between-construct growth factors. 

These MGMs have been shown useful in examining joint nonlinear development; however, the relationship between multiple longitudinal outcomes may also be unidirectional in multiple domains \citep{Liu2021PBLSGM}. In the educational field, for example, reading is a crucial developmental process with wide-ranging implications for later academic performance \citep{Torgesen2002Reading}. Accordingly, the investigation of how the early development of reading affects the later development of other disciplinary subjects such as mathematics and science may be more important than the alternative since \citet{Akba2016Reading} has shown that reading comprehension is one of the most critical factors that contribute to mathematics or science achievement. Similarly, in the biomedical area, we often first observe treatment effects from clinical indices and then from patient-reported outcomes (PROs) such as measures related to illness burden or health-related quality of life. In such cases, it is of greater interest to examine how the change in clinical features affects the change in the PROs. The present study aims to develop longitudinal models to investigate mediational processes to estimate regression coefficients between growth factors.

\subsection*{Simple Mediation Model}\label{I:Cross}
In this section, we first introduce a simple mediation model, which is represented in the diagram by Figure \ref{fig:Med} \citep{Baron1986indirect}. As shown in the figure, a simple mediation model contains three variables, a predictor $X$, a mediator $M$, and an outcome $Y$, with $X$ is proposed as influencing $Y$ through $M$. Note that in this setup, both $M$ and $Y$ are dependent variables. There are two possible pathways through which the predictor $X$ can impact $Y$: the path leads from $X$ to $Y$ without passing through $M$ (referred to as the direct effect of $X$ on $Y$), and the path from $X$ to $Y$ through $M$ (referred to as the indirect effect or mediated effect of $X$ on $Y$). The direct effect represents how $Y$ is influenced by $X$ directly, while the indirect effect represents how $Y$ is affected by $X$ through a causal sequence where $X$ affects $M$, which, in turn, impacts $Y$. 

\figurehere{1}

The primary research question when employing a mediation model is estimating and interpreting the direct and indirect effects. For example, in the simple mediation model, the estimation of these effects is through two linear regression models, one for each dependent variable, as shown below
\begin{equation}\nonumber
\begin{aligned}
&M=\beta_{M0}+a\times X+\epsilon_{M},\\
&Y=\beta_{Y0}+c\times X+b\times M+\epsilon_{Y},
\end{aligned}
\end{equation}
where $\beta_{M0}$ and $\beta_{Y0}$ are regression constants,  $\epsilon_{M}$ and $\epsilon_{Y}$ are errors of $M$ and $Y$, respectively. In addition, $a$, $b$, and $c$ are regression coefficients, among which $c$ quantifies the direct effect of $X$ on $Y$, while the product of the other two coefficients, $a\times b$, quantifies the indirect effect of $X$ on $Y$. According to \citet{Baron1986indirect}, the most common situation in social science research is partial mediation, where the mediator only explains part of the relationship between the predictor and the outcome. Therefore, the total effect of $X$ on $Y$ is the sum of direct and indirect effects $c+a\times b$ \citep{MacKinnon2000Mediator}.

One challenge in the mediation analyses is how to obtain standard errors of mediated effects, based on which we can conduct a hypothesis test for statistically significant partial mediation and construct a $95\%$ confidence interval (CI) for the indirect effect. Generally, earlier studies have recommended employing the Delta method, for example, \citet{Sobel1982indirect, Sobel1986indirect, MacKinnon2002Mediator} to approximate the standard error of the product of $a$ and $b$, or utilizing bootstrap techniques to calculate more accurate estimates \citep{Hayes2009Mediate, Shrout2002Mediate}. Moreover, \citet{Cheung2008Mediate} has shown that the bias-corrected bootstrap CI performs best in the tests for the mediated effects compared to other methods. In the present study, we do not intend to develop a novel approach to obtaining standard errors of mediated effects or compare different approaches for constructing $95\%$ CIs. Instead, we utilize the Delta method to obtain standard errors of mediated effects as more accurate standard errors are out of the primary research interests of the current study. Note that the standard errors calculated using the Delta method can be automatically obtained from multiple software of the structural equation modeling (SEM) framework such as \textit{Mplus} 8 \citep{Muthen2017Mplus} and \textit{OpenMx} \citep{User2020OpenMx}.

Earlier studies, for example, \citet{Gollob1987mediate} have described multiple fundamental problems of applying mediation models to cross-sectional data. On the one hand, the causal relationships often take time to unfold, implying that the magnitude of the causal effect usually depends on the elapsed time between measurement occasions of the predictor, mediator, and outcome. However, the utilization of cross-sectional data assumes that the causal effects are instantaneous and that the magnitude of a causal effect remains the same over time. On the other hand, using cross-sectional data leaves out measurements at previous times that are potential predictors of the constructed model. Accordingly, longitudinal data are favored for testing the mediation hypotheses. Earlier studies proposed applying the simple mediation model to analyze data where the predictor, mediator, and outcome are from multiple waves. Such applications focus on interindividual differences. However, a critical aspect of a longitudinal study is intraindividual change, which cannot be analyzed explicitly by the simple mediation model \citep{Selig2009mediate}. Additionally, \citet{Cole2003mediate, Maxwell2007mediate} have demonstrated that the simple mediation model typically yields substantially biased estimates for longitudinal parameters. 

\subsection*{Longitudinal Mediation Model}\label{I:Longitudinal}
In this section, we briefly review modeling frameworks to investigate mediation in longitudinal data, which allow for the examination of intraindividual change. Researchers have explored mediational processes in three frameworks: cross-lagged panel models \citep{Cole2003mediate}, latent growth curve models \citep{Cheong2003mediate}, and latent change score models \citep[Chapter~8]{MacKinnon2008mediate}. \citet{Cole2003mediate} and \citet[Chapter~8]{MacKinnon2008mediate} have provided extensive overviews of the application of cross-lagged panel mediation models. At least three waves of all three constructs in a fully-specified model, the predictor, the mediator, and the outcome are needed to test how the predictor at the first wave (i.e., $t_{1}$) affects the mediator at the second wave (i.e., $t_{2}$), and in turn, influences the outcome at the third wave (i.e., $t_{3}$).

The second type of longitudinal mediation model stems from the latent growth curve modeling framework. In the scenario where there are a baseline predictor (i.e., a time-invariant predictor), a longitudinal mediator, and a longitudinal outcome, researchers employ linear growth curve models to depict the intraindividual change of the mediator and the outcome and examine how a baseline predictor affects the status or change of the mediator affects the change in the outcome. For example, \citet{Cheong2003mediate} describes how an intervention program (the predictor) affects the change in the outcome through the change of the mediator using a parallel latent growth curve model. \citet{Soest2011mediate} showed that, other than the change in the longitudinal mediator process, its initial status can also be viewed as a mediator in a longitudinal mediation model with a note that the regression of the initial status is comparable to the cross-sectional mediation analysis. In addition, \citet[Chapter~8]{MacKinnon2008mediate} added that the predictor can also be a longitudinal variable and demonstrated how to extend the model to perform mediation analyses among three parallel longitudinal processes. 

Earlier studies such as \citet[Chapter~8]{MacKinnon2008mediate} and \citet{Selig2009mediate} also extended the latent change score modeling framework \citep{McArdle1994LCSM}, where intraindividual change is captured by latent change scores, to examine longitudinal mediation. This modeling framework allows for the examination of mediated effects among the status of the predictor, mediator, and outcome. Its additional advantage is that it enables researchers to investigate how the change score of the predictor during the first time interval (i.e., $t_{2}-t_{1}$) influences the change score of the mediator during the second time interval (i.e., $t_{3}-t_{2}$), and in turn, affects the change score of the outcome during the third time interval (i.e., $t_{4}-t_{3}$). \citet{Selig2009mediate} have shown that the mediation effects can also be any combination of statuses and change scores of constructs. Earlier studies, for example, \citet[Chapter~8]{MacKinnon2008mediate} and \citet{Selig2009mediate}, have provided a detailed discussion in terms of the theory of change for the three variables (i.e., the predictor, mediator, and outcome), the role of time, and the types of indirect effects of these three longitudinal mediation models.

The present study focuses on the second type of longitudinal mediation model, latent growth mediation models. The existing latent growth mediation models developed in \citet{Cheong2003mediate} and \citet[Chapter~8]{MacKinnon2008mediate} assume that change patterns of each repeated variable take the linear functional form. Given that a process may show nonlinear change patterns with respect to time, \citet{Cheong2003mediate} suggested employing a two-stage piecewise (also referred to as a bilinear spline, see \citet[Chapter~12]{Grimm2016growth}) parallel growth model instead of the parallel linear growth model to evaluate nonlinear trajectories. Accordingly, in this present study, we developed two models to investigate mediational processes where the parallel bilinear spline growth model (PBLSGM) \citep{Liu2021PBLSGM} is employed to examine joint longitudinal processes. The first model can be utilized to evaluate a mediational process with a baseline predictor and subsequently two variables that are repeatedly measured over time, while the second model can be employed to investigate a mediational process with three longitudinal variables. 

There are two statistical challenges to utilizing a linear-linear piecewise functional form. First, the change point or knot where the change of the rate-of-change occurs must be determined. Earlier studies have demonstrated that the knot can either be pre-specified by domain theories \citep{Dumenci2019knee, Flora2008knot, Riddle2015knee, Sterba2014individually} or be estimated as an unknown parameter if the knowledge of the knot is unavailable or the estimation of the knot itself is of research interest. \citep{Cudeck2003knot_F, Harring2006nonlinear, Kwok2010simu, Kohli2011PLGC, Kohli2013PLGC1, Kohli2013PLGC2, Preacher2015repara, Liu2019knot, Liu2019BLSGM, Liu2019BLSGMM, Dominicus2008knot_B, McArdle2008knot_B, Wang2008knot_B, Muniz2011knot_B, Kohli2015PLGC1, Lock2018knot_B}. In the two proposed models, we assume that the knot of each longitudinal process is unknown and to be estimated for two considerations. On the one hand, time is important in a longitudinal mediation model, and for each process, the knot is the transition time of the two stages. Accordingly, we want to avoid any unnecessary pre-specification of the knot. On the other hand, the knot can be viewed as a developmental milestone or an event in a longitudinal process. For example, when examining sleep behavior of infants over time, the knot could be the age of crawling since the sleep behavior is expected to be different pre- and post- crawling age \citep[Chapter~10]{Grimm2016growth}. Similarly, when analyzing the recovery process from knee replacement surgery, the knot is viewed as the transition time from the recovery of surgical pain to the gradual recovery period \citep{Dumenci2019knee}. So ideally, the knot of the predictor process occurs first, followed by the knot of the mediator process, and then the knot of the outcome process. We want to obtain data-driven evidence for this temporal order from the proposed models.

Second, for a longitudinal process that takes a linear-linear piecewise functional form, we have three statuses, the initial status, the status at the knot (i.e., the change-point at which two linear segments join together), and the status at the end of the study. In addition, we have a slope for each of the two linear pieces, which captures the short-term change rate and the long-term change rate, respectively. \citet{Harring2006nonlinear} has shown that the degrees of freedom of the linear-linear piecewise functional form is four; therefore, we cannot freely estimate all the statuses and slopes. Most existing studies with an unknown knot usually view the initial status, two slopes, and the knot as four free growth coefficients \citep{Harring2006nonlinear, Preacher2015repara, Liu2019BLSGM, Peralta2020PBLSGM, Liu2021PBLSGM}. In this project, we consider the two slopes, the knot and the measurement at the knot, as the free growth coefficients since the measurement at the knot is more relevant than the initial status to longitudinal mediation analysis given two reasons. First, the initial status of a longitudinal variable only provides information at baseline; therefore, the regression of the initial status of the longitudinal mediator on that of the predictor only estimates instantaneous effects, and so does the regression of the initial status of the longitudinal outcome \citep{Soest2011mediate}. Second, as stated earlier, the knot is usually viewed as a milestone or an event in a longitudinal process; therefore, the status of the knot in one longitudinal variable may influence a long-term change in the other longitudinal process. Suppose we have two types of endpoints to evaluate the treatment effects of psychotherapy: one being a clinical indicator and the other a patient-reported outcome (PRO) related to illness burden and health-related quality of life. It is often expected to observe improvement in clinical indicators first and arrive at a plateau, leading to improved PROs. Therefore, the measurement at the plateau (i.e., the knot) of a clinical indicator is likely to affect the improvement in PROs later on.

Based on the above considerations, we view the knot as unknown for each longitudinal process in the two proposed models and consider the two slopes, the knot, and the status at the knot as the free growth coefficients. We then estimate the two slopes and the status of the knot at the individual level with an assumption that each process-specified knot is fixed across all individuals\footnote{Although multiple existing studies, for example, \citet{Preacher2015repara, Liu2019BLSGM, Peralta2020PBLSGM, Liu2021PBLSGM} have demonstrated that the variance of the knot can also be estimated, we assume that each process-specified knot is fixed across all individuals since the knot variance is out of the research interests of the present study. }. Additionally, similar to \citet{Liu2021PBLSGM}, we construct the proposed models in the framework of individual measurement occasions using the `definition variables' approach \citep{Mehta2000people, Mehta2005people, Sterba2014individually} to avoid potential inadmissible estimation \citep{Blozis2008coding, Coulombe2015ignoring} and allow for different time structures across constructs \citep{Liu2021PBLSGM}. 

The remainder of this article is organized as follows. In the method section, we start from a bilinear spline growth model to estimate a fixed knot for a univariate longitudinal process. We then extend it to longitudinal mediation models and introduce the model specification of the proposed models. We define the mediational process as either the baseline covariate or the change in covariate influencing the change in the mediator, which, in turn, affects the change in the outcome. Next, we describe the model estimation and model evaluation that is realized by the Monto Carlo simulation. We then present simulation results and evaluate the proposed models in terms of non-convergence rate and the performance measures, which include the relative bias, the empirical standard error (SE), the relative root-mean-squared error (RSME), and the empirical coverage probability for a nominal $95\%$ confidence interval of each parameter. Next, in the application section, we demonstrate how to apply the proposed models to examine a data set of longitudinal records of reading, mathematics, and science abilities from ECLS-K: $2011$ \citep{Tourangeau2013ECLS}. Finally, discussions are framed regarding practical considerations, methodological considerations, and future directions.

\section*{Method}\label{Method}
\subsection*{Bilinear Spline Growth Curve Model with a Fixed Knot}\label{M:Uni}
In this section, we briefly describe a latent growth curve (LGC) model with a linear-linear functional form to estimate a fixed knot in the framework of individual measurement occasions. As shown in Figure \ref{fig:knot}, we specify a separate linear function for each of the two phases of the developmental process and express the measurement for the $i^{th}$ individual at the $j^{th}$ time in the framework of individual measurement occasions as
\begin{equation}\label{eq:seg}
y_{ij}=\begin{cases}
\eta^{[y]}_{0i}+\eta^{[y]}_{1i}t_{ij}+\epsilon^{[y]}_{ij} & t_{ij}\le\gamma^{[y]}\\
\eta^{[y]}_{0i}+\eta^{[y]}_{1i}\gamma+\eta^{[y]}_{2i}(t_{ij}-\gamma^{[y]})+\epsilon^{[y]}_{ij} & t_{ij}>\gamma^{[y]}\\
\end{cases},
\end{equation}
where $y_{ij}$ and $t_{ij}$ are the measurement and measurement occasion of the $i^{th}$ individual at wave $j$. In Equation (\ref{eq:seg}), $\eta^{[y]}_{0i}$, $\eta^{[y]}_{1i}$ and $\eta^{[y]}_{2i}$ are the individual-level intercept, first slope and second slope, which are usually called `growth factors' in the latent growth curve modeling framework; they three along with the fixed knot $\gamma^{[y]}$ together determine the functional form of the growth curve of $\boldsymbol{y}_{i}$.

\figurehere{2}

Existing SEM software such as \textit{Mplus} \citep{Muthen2017Mplus} and the \textit{R} package \textit{OpenMx} \citep{User2020OpenMx} does not allow for a conditional statement of an unknown parameter such as $\gamma^{[y]}$ in Equation (\ref{eq:seg}), so the piecewise functional form above cannot be specified directly. In order to unify pre- and post-knot expressions, we have to reparameterize growth factors, which can be realized through multiple approaches. 
\citet{Harring2006nonlinear} proposed to reparameterize three growth factors in Equation (\ref{eq:seg}) (i.e., $\eta^{[y]}_{0i}$, $\eta^{[y]}_{1i}$ and $\eta^{[y]}_{2i}$) to the average of the two intercepts, the average of the two slopes, and the half difference between the two slopes. Alternatively, \citet[Chapter~11]{Grimm2016growth} suggested reexpressing the three growth factors to the measurement at the knot and two slopes. Additionally, \citet{Liu2019BLSGM} reparameterized $\eta^{[y]}_{0i}$, $\eta^{[y]}_{1i}$ and $\eta^{[y]}_{2i}$ as the measurement at the knot, the average of the two slopes, and the half difference between the two slopes. 

The reparameterized growth factors proposed in \citet{Harring2006nonlinear, Liu2019BLSGM} are no longer directly related to the underlying developmental process and therefore are less useful in the present study where we want to estimate regression coefficients between original growth factors. Accordingly, we follow the reparameterized approach in \citet[Chapter~11]{Grimm2016growth}, where all three reparameterized coefficients are still directly related to the growth patterns. We then write the repeated outcome as
\begin{align}
y_{ij}
&=(\eta^{[y]}_{0i}+\eta^{[y]}_{1i}\gamma^{[y]})+\eta^{[y]}_{1i}\min(0,t_{ij}-\gamma^{[y]})+\eta^{[y]}_{2i}\max(0,t_{ij}-\gamma^{[y]})+\epsilon^{[y]}_{ij}\nonumber\\
&=\eta^{[y]}_{\gamma_{i}}+\eta^{[y]}_{1i}\min(0,t_{ij}-\gamma^{[y]})+\eta^{[y]}_{2i}\max(0,t_{ij}-\gamma^{[y]})+\epsilon^{[y]}_{ij}\nonumber,
\end{align}
where $\eta^{[y]}_{\gamma_{i}}$ is the measurement at the knot of the $i^{th}$ individual. 

\subsection*{Model Specification of Mediation Model with Baseline Predictor, Longitudinal Mediator, and Longitudinal Outcome}\label{M:Bi}
In this section, we extend the univariate linear-linear latent growth curve model to analyze a mediational process with baseline predictor, longitudinal mediator, and longitudinal outcome. Suppose both the mediator and the outcome take the linear-linear functional form with an unknown fixed knot. In such situations, we can construct a model for a mediational process where the baseline predictor influences the change in the outcome directly and indirectly through its effect on the change in the mediator. The longitudinal mediation model can be specified as
\begin{equation}\label{eq:bi_Med}
\begin{pmatrix}
\boldsymbol{m}_{i} \\ \boldsymbol{y}_{i}
\end{pmatrix}=
\begin{pmatrix}
\boldsymbol{\Lambda}_{i}^{[m]} & \boldsymbol{0} \\ \boldsymbol{0} & \boldsymbol{\Lambda}_{i}^{[y]}
\end{pmatrix}\times
\begin{pmatrix}
\boldsymbol{\eta}^{[m]}_{i} \\ \boldsymbol{\eta}^{[y]}_{i}
\end{pmatrix}+
\begin{pmatrix}
\boldsymbol{\epsilon}^{[m]}_{i} \\ \boldsymbol{\epsilon}^{[y]}_{i}
\end{pmatrix},
\end{equation}
where $\boldsymbol{m}_{i}$ and $\boldsymbol{y}_{i}$ are a $J\times1$ vector of the repeated measurements of the mediator and outcome of the $i^{th}$ individual, respectively (in which $J$ is the number of measurement occasions). With an assumption that the trajectories of mediator and outcome process take the linear-linear functional form with an unknown knot, $\boldsymbol{\eta}^{[u]}_{i}$ ($u=m, y$) is a $3\times1$ vector of growth factors,
\begin{equation}\nonumber
\boldsymbol{\eta}^{[u]}_{i} = \left(\begin{array}{rrr}
\eta^{[u]}_{1i} & \eta^{[u]}_{\gamma_{i}} & \eta^{[u]}_{2i} 
\end{array}\right)^{T},
\end{equation}
in which $\eta^{[u]}_{1i}$, $\eta^{[u]}_{\gamma_{i}}$, and $\eta^{[u]}_{2i}$ represent the slope of the first stage, the measurement at the knot, and the slope of the second stage, respectively. The corresponding factor loadings $\boldsymbol{\Lambda}^{[u]}_{i}$, a $J\times3$ matrix, is expressed as
\begin{equation}\nonumber
\begin{aligned}
&\boldsymbol{\Lambda}^{[u]}_{i} = \left(\begin{array}{rrr}
\min(0,t_{ij}-\gamma^{[u]}) & 1 & \max(0,t_{ij}-\gamma^{[u]})
\end{array}\right)
&(j=1,\cdots, J).
\end{aligned}
\end{equation}
where the subscript $i$ indicates that we build the model in the framework of individual measurement occasions. Additionally, $\boldsymbol{\epsilon}^{[u]}_{i}$ is a $J\times1$ vector of residuals of the $i^{th}$ individual.

To define a mediation model with a baseline predictor, a longitudinal mediator, and a longitudinal outcome, we regress the growth factors of the mediator process on the predictor for the $i^{th}$ individual
\begin{equation}\label{eq:M_reg1}
\boldsymbol{\eta}^{[m]}_{i}=\boldsymbol{\alpha}^{[m]}+\boldsymbol{B}^{[x\rightarrow{m}]}\times x_{i}+\boldsymbol{\zeta}^{[m]}_{i},
\end{equation}
where $x_{i}$, which is either continuous or binary, is the baseline covariate of the $i^{th}$ individual, $\boldsymbol{\alpha}^{[m]}$ is a $3\times1$ vector of growth factor intercepts of the mediator process (which is the mean vector of growth factors of the mediator if the covariate is centered), and $\boldsymbol{B}^{[x\rightarrow{m}]}$ is a $3\times1$ vector of regression coefficients from the predictor to the first slope, to the measurement at the knot, and to the second slope of the mediator process, that is
\begin{equation}\label{eq:Beta_xm1}
\boldsymbol{B}^{[x\rightarrow{m}]}=\left(\begin{array}{rrr}
\beta^{[x\rightarrow{m}]}_{1} & \beta^{[x\rightarrow{m}]}_{\gamma} & \beta^{[x\rightarrow{m}]}_{2}
\end{array}\right)^{T}.
\end{equation}
Additionally, we need to regress the growth factors of the outcome process on the predictor and the growth factors of the mediator process
\begin{equation}\label{eq:Y_reg1}
\boldsymbol{\eta}^{[y]}_{i}=\boldsymbol{\alpha}^{[y]}+\boldsymbol{B}^{[x\rightarrow{y}]}\times x_{i}+\boldsymbol{B}^{[m\rightarrow{y}]}\times\boldsymbol{\eta}^{[m]}_{i}+\boldsymbol{\zeta}^{[y]}_{i},
\end{equation}
where $\boldsymbol{\alpha}^{[y]}$ is a $3\times1$ vector of growth factor intercepts of the outcome process, $\boldsymbol{B}^{[x\rightarrow{y}]}$ is a $3\times1$ vector of regression coefficients from the predictor to the first slope, to the measurement at the knot, and to the second slope of the outcome process, that is
\begin{equation}\label{eq:Beta_xy1}
\boldsymbol{B}^{[x\rightarrow{y}]}=\left(\begin{array}{rrr}
\beta^{[x\rightarrow{y}]}_{1} & \beta^{[x\rightarrow{y}]}_{\gamma} & \beta^{[x\rightarrow{y}]}_{2}
\end{array}\right)^{T}.
\end{equation}
Additionally, $\boldsymbol{B}^{[m\rightarrow{y}]}$ is a $3\times3$ matrix of regression coefficients from the growth factors of the mediator process to those of the outcome process. The coefficient matrix can be expressed as
\begin{equation}\label{eq:Beta_my}
\boldsymbol{B}^{[m\rightarrow{y}]}=\left(\begin{array}{rrr}
\beta^{[m\rightarrow{y}]}_{11} & 0 & 0 \\
\beta^{[m\rightarrow{y}]}_{1\gamma} & \beta^{[m\rightarrow{y}]}_{\gamma\gamma} & 0 \\
\beta^{[m\rightarrow{y}]}_{12} & \beta^{[m\rightarrow{y}]}_{\gamma2} & \beta^{[m\rightarrow{y}]}_{22} \\
\end{array}\right).
\end{equation}
In Equations (\ref{eq:Beta_xm1}), (\ref{eq:Beta_xy1}), (\ref{eq:Beta_my}),
the superscript and subscript of $\beta$ together define the path of the corresponding coefficient. For example, $\beta^{[x\rightarrow{m}]}_{1}$ is the path coefficient from the baseline covariate to the first slope of the mediator process. Similarly, $\beta^{[m\rightarrow{y}]}_{1\gamma}$ is the coefficient from the first slope of the mediator process to the knot measurement of the outcome process. Additionally, $\boldsymbol{\zeta}^{[u]}_{i}$ is a $3\times1$ vector of deviations of the $i^{th}$ individual from the mean values of the variable-specific growth factors. We list six possible paths of indirect effects of the predictor on the outcome process in the specified model in Table \ref{tbl:path}.

\tablehere{1}

\subsection*{Model Specification of Mediation Model with Longitudinal Predictor, Mediator and Outcome}\label{M:Tri}
In this section, we extend the univariate linear-linear latent growth curve model to investigate a mediational process with a longitudinal predictor, mediator, and outcome. Suppose all three variables take a linear-linear functional form with an unknown fixed knot. In such situations, we can construct a longitudinal mediation model to explore how the change in covariate influences the change in the outcome directly and indirectly through its effect on the change in the mediator. The longitudinal mediation model can be specified as
\begin{equation}\label{eq:tri_Med}
\begin{pmatrix}
\boldsymbol{x}_{i} \\ \boldsymbol{m}_{i} \\ \boldsymbol{y}_{i}
\end{pmatrix}=
\begin{pmatrix}
\boldsymbol{\Lambda}_{i}^{[x]} & \boldsymbol{0} & \boldsymbol{0} \\
\boldsymbol{0} & \boldsymbol{\Lambda}_{i}^{[m]} & \boldsymbol{0} \\ 
\boldsymbol{0} & \boldsymbol{0} & \boldsymbol{\Lambda}_{i}^{[y]}
\end{pmatrix}\times
\begin{pmatrix}
\boldsymbol{\eta}^{[x]}_{i} \\\boldsymbol{\eta}^{[m]}_{i} \\ \boldsymbol{\eta}^{[y]}_{i}
\end{pmatrix}+
\begin{pmatrix}
\boldsymbol{\epsilon}^{[x]}_{i} \\\boldsymbol{\epsilon}^{[m]}_{i} \\ \boldsymbol{\epsilon}^{[y]}_{i}
\end{pmatrix}.
\end{equation}
which defines a longitudinal process for the predictor in addition to the mediator and outcome processes specified in Equation (\ref{eq:bi_Med}). In Equation (\ref{eq:tri_Med}), $\boldsymbol{x}_{i}$ is a $J\times1$ vector of the repeated measures of the predictor for the $i^{th}$ individual. In addition, $\boldsymbol{\eta}^{[x]}_{i}$, $\boldsymbol{\Lambda}^{[x]}_{i}$, and  $\boldsymbol{\epsilon}^{[x]}_{i}$ are the growth factors (a $3\times1$ vector), the corresponding factor loadings (a $J\times3$ matrix), and the residuals (a $J\times1$ vector) of the covariate process, respectively. We then write the growth factors of the covariate process as deviations from the corresponding growth factor means
\begin{equation}\label{eq:X_reg2}
\boldsymbol{\eta}^{[x]}_{i}=\boldsymbol{\mu}^{[x]}_{\boldsymbol{\eta}}+\boldsymbol{\zeta}^{[x]}_{i},
\end{equation}
where $\boldsymbol{\mu}^{[x]}_{\boldsymbol{\eta}}$ and $\boldsymbol{\zeta}^{[x]}_{i}$ are the mean vector of growth factors (a $3\times1$ vector) and the deviations (a $3\times1$ vector) of the covariate process. The growth factors of the mediator process are regressed on the those of the covariate, as shown below
\begin{equation}\label{eq:M_reg2} 
\boldsymbol{\eta}^{[m]}_{i}=\boldsymbol{\alpha}^{[m]}+\boldsymbol{B}^{[x\rightarrow{m}]}\times\boldsymbol{\eta}^{[x]}_{i}+\boldsymbol{\zeta}^{[m]}_{i},
\end{equation}
where $\boldsymbol{\alpha}^{[m]}$ is a $3\times1$ vector of growth factor intercepts of the mediator, $\boldsymbol{B}^{[x\rightarrow{m}]}$ is a $3\times3$ matrix of regression coefficients from the growth factors of the covariate process to those of the mediator process
\begin{equation}\label{eq:Beta_xm2}
\boldsymbol{B}^{[x\rightarrow{m}]}=\left(\begin{array}{rrr}
\beta^{[x\rightarrow{m}]}_{11} & 0 & 0 \\
\beta^{[x\rightarrow{m}]}_{1\gamma} & \beta^{[x\rightarrow{m}]}_{\gamma\gamma} & 0 \\
\beta^{[x\rightarrow{m}]}_{12} & \beta^{[x\rightarrow{m}]}_{\gamma2} & \beta^{[x\rightarrow{m}]}_{22} \\
\end{array}\right).
\end{equation}
Similarly, the growth factors of the outcome process are regressed on those of the covariate and mediator
\begin{equation}\label{eq:Y_reg2}
\boldsymbol{\eta}^{[y]}_{i}=\boldsymbol{\alpha}^{[y]}+\boldsymbol{B}^{[x\rightarrow{y}]}\times\boldsymbol{\eta}^{[x]}_{i}+\boldsymbol{B}^{[m\rightarrow{y}]}\times\boldsymbol{\eta}^{[m]}_{i}+\boldsymbol{\zeta}^{[y]}_{i},
\end{equation}
where $\boldsymbol{\alpha}^{[y]}$ is a $3\times1$ vector of growth factor intercepts of the outcome, $\boldsymbol{B}^{[x\rightarrow{y}]}$ ($\boldsymbol{B}^{[m\rightarrow{y}]}$) is a $3\times3$ matrix of regression coefficients from the growth factors of the covariate (mediator) process to those of the outcome process. The coefficient matrix $\boldsymbol{B}^{[x\rightarrow{y}]}$ is defined as 
\begin{equation}\label{eq:Beta_xy2}
\boldsymbol{B}^{[x\rightarrow{y}]}=\left(\begin{array}{rrr}
\beta^{[x\rightarrow{y}]}_{11} & 0 & 0 \\
\beta^{[x\rightarrow{y}]}_{1\gamma} & \beta^{[x\rightarrow{y}]}_{\gamma\gamma} & 0 \\
\beta^{[x\rightarrow{y}]}_{12} & \beta^{[x\rightarrow{y}]}_{\gamma2} & \beta^{[x\rightarrow{y}]}_{22} \\
\end{array}\right)
\end{equation}
and $\boldsymbol{B}^{[m\rightarrow{y}]}$ has the same expression as such in Equation (\ref{eq:Beta_my}). Ten possible paths of indirect effects of the predictor process on the outcome process in the specified model are also listed in Table \ref{tbl:path}.

\subsection*{Model Estimation}\label{M:est}
To simplify estimation, we make the following four assumptions. First, we assume that the covariate's growth factors are normally distributed. Second, we assume that the mediator's growth factors are normally distributed conditional on the baseline predictor (or growth factors of the predictor process). The third assumption is that the outcome's growth factors are normally distributed conditional on the baseline covariate (or growth factors of the covariate process) and the growth factors of the mediator process. Accordingly, $\boldsymbol{\zeta}^{[u]}_{i}\sim\text{MVN}(\boldsymbol{0}, \boldsymbol{\Psi}^{[u]}_{\boldsymbol{\eta}})$ ($u=x,m,y$), where $\boldsymbol{\Psi}^{[u]}_{\boldsymbol{\eta}}$ is a $3\times3$ (unexplained) variance-covariance matrix of growth factors of the variable \textit{u}. We also assume that the individual residuals $\boldsymbol{\epsilon}^{[u]}_{i}$ are identical and independent normal distributions over time, and the covariances between residuals are homogeneous over time. We define $\boldsymbol{I}$ as a $J\times J$ identity matrix, and for the first and second proposed models, the corresponding variance-covariance matrix of residuals can be expressed as 
\begin{equation}\nonumber
\begin{pmatrix} 
\boldsymbol{\epsilon}^{[m]}_{i} \\ \boldsymbol{\epsilon}^{[y]}_{i}
\end{pmatrix}\sim \text{MVN}\bigg(\boldsymbol{0}, 
\begin{pmatrix}
\theta^{[m]}_{\epsilon}\boldsymbol{I} & \theta^{[my]}_{\epsilon}\boldsymbol{I} \\
& \theta^{[y]}_{\epsilon}\boldsymbol{I}
\end{pmatrix}\bigg),
\end{equation}
and
\begin{equation}\nonumber
\begin{pmatrix} 
\boldsymbol{\epsilon}^{[x]}_{i} \\ \boldsymbol{\epsilon}^{[m]}_{i} \\ \boldsymbol{\epsilon}^{[y]}_{i}
\end{pmatrix}\sim \text{MVN}\bigg(\boldsymbol{0}, 
\begin{pmatrix}
\theta^{[x]}_{\epsilon}\boldsymbol{I} & \theta^{[xm]}_{\epsilon}\boldsymbol{I} & \theta^{[xy]}_{\epsilon}\boldsymbol{I}\\
& \theta^{[m]}_{\epsilon}\boldsymbol{I} & \theta^{[my]}_{\epsilon}\boldsymbol{I} \\
& & \theta^{[y]}_{\epsilon}\boldsymbol{I}
\end{pmatrix}\bigg),
\end{equation}
respectively.

The parameters in the first proposed model include the mean and variance of the covariate, the intercepts ($\boldsymbol{\alpha}^{[u]}$), and unexplained variance-covariance ($\boldsymbol{\Psi}^{[u]}_{\boldsymbol{\eta}}$) of the growth factors of the mediator (outcome) process, the coefficients from the covariate to the mediator (outcome) growth factors and those from the mediator growth factors to the outcome growth factors,  the mediator (outcome) residual variance, and the residual covariance. We define $\boldsymbol{\Theta}_{1}$ as
\begin{equation}\nonumber
\boldsymbol{\Theta}_{1}=\{\mu_{x}, \Phi_{x}, \boldsymbol{\alpha}^{[u]}, \boldsymbol{B}^{[x\rightarrow{m}]}, \boldsymbol{B}^{[x\rightarrow{y}]}, \boldsymbol{B}^{[m\rightarrow{y}]}, \boldsymbol{\Psi}^{[u]}_{\boldsymbol{\eta}}, \theta^{[u]}_{\epsilon}, \theta^{[my]}_{\epsilon}\}\ (u=m, y)
\end{equation}
to list the parameters specified in the first proposed longitudinal mediation model, where $\boldsymbol{B}^{[x\rightarrow{m}]}$ and $\boldsymbol{B}^{[x\rightarrow{y}]}$ are those defined in Equations (\ref{eq:Beta_xm1}) and (\ref{eq:Beta_xy1}), respectively.

The parameters in the second proposed model include the mean vector ($\boldsymbol{\mu}^{[x]}_{\boldsymbol{\eta}}$) and variance-covariance matrix ($\boldsymbol{\Psi}^{[x]}_{\boldsymbol{\eta}}$) of the growth factors of the predictor process, the intercepts ($\boldsymbol{\alpha}^{[u]}$) and unexplained variance-covariance ($\boldsymbol{\Psi}^{[u]}_{\boldsymbol{\eta}}$) of the growth factors of the mediator (outcome) process, the coefficients between growth factors of three longitudinal processes, the residual variances and covariances. We define $\boldsymbol{\Theta}_{2}$ as
\begin{equation}\nonumber
\boldsymbol{\Theta}_{2}=\{\boldsymbol{\mu}^{[x]}_{\boldsymbol{\eta}}, \boldsymbol{\Psi}^{[x]}_{\boldsymbol{\eta}}, \boldsymbol{\alpha}^{[u]}, \boldsymbol{B}^{[x\rightarrow{m}]}, \boldsymbol{B}^{[x\rightarrow{y}]}, \boldsymbol{B}^{[m\rightarrow{y}]}, \boldsymbol{\Psi}^{[u]}_{\boldsymbol{\eta}}, \theta^{[x]}_{\epsilon}, \theta^{[u]}_{\epsilon}, \theta^{[xm]}_{\epsilon}, \theta^{[xy]}_{\epsilon}, \theta^{[my]}_{\epsilon}\}\ (u=m, y)
\end{equation}
and list the parameters specified in the second longitudinal mediation model, where $\boldsymbol{B}^{[x\rightarrow{m}]}$ and $\boldsymbol{B}^{[x\rightarrow{y}]}$ are those defined in Equations (\ref{eq:Beta_xm2}) and (\ref{eq:Beta_xy2}), respectively.

We estimate $\boldsymbol{\Theta}_{1}$ and $\boldsymbol{\Theta}_{2}$ using full information maximum likelihood (FIML) to account for the individual measurement occasions and the potential heterogeneity of individual contributions to the likelihood function. In this work, we constructed the proposed longitudinal mediation models using the R package \textit{OpenMx} with CSOLNP optimizer \citep{Pritikin2015OpenMx, OpenMx2016package, User2020OpenMx, Hunter2018OpenMx}. 

In addition to the parameters that can be estimated directly, in practice, we are also interested in estimating the conditional mean vector and variance-covariance matrix of the growth factors of the mediator (outcome) process and indirect effects as well as the total effects listed in Table \ref{tbl:path}. When fitting the model using the \textit{R} package \textit{OpenMx}, we can specify the additional parameters in the function \textit{mxAlgebra()} \citep{User2020OpenMx}. We need to provide the algebraic expressions of the conditional mean vector and variance-covariance matrix of growth factors of the mediator (outcome) process and those in Table \ref{tbl:path}; then \textit{OpenMx} is capable of computing the point estimates along with their standard errors of these additional parameters (that is realized by the Delta Method). We provide the expressions of the conditional mean vector and variance-covariance matrix of growth factors of the mediator (outcome) process for the two models in Appendix \ref{supp:1} and Appendix \ref{supp:2}, respectively. Additionally, we provide \textit{OpenMx} code in the online appendix (\url{https://github.com/Veronica0206/Extension_projects}) and a demonstration to show how to build the proposed models and estimate these additional parameters. 

\section*{Model Evaluation}\label{Evaluation}
In this section, we perform simulation studies to evaluate the performance of the proposed models. We first discuss the simulation design for the two longitudinal mediation models, along with the performance metrics and the consideration for the number of replications. After that, we describe the general steps for data generation.

\subsection*{Design of Simulation Study}\label{E:design}
We provide the detailed simulation designs for the two longitudinal mediation models in Table S1 and Table S2. The parameters of the most interest in the proposed model are the coefficients $\boldsymbol{B}^{[x\rightarrow{m}]}$, $\boldsymbol{B}^{[x\rightarrow{y}]}$ and $\boldsymbol{B}^{[m\rightarrow{y}]}$, based on which we obtain direct, indirect, and total effects that the baseline predictor (or predictor process) has on the outcome process. The conditions hypothesized to influence the estimation of these coefficients and other model parameters include sample size, the number of repeated measurements, the knot locations, the ratio of indirect effects to direct effects, shapes of trajectories, and measurement precision. Accordingly, we fixed the factors, for example, the mean vectors and variance-covariance matrices of the growth factors of longitudinal processes, that presumably do not affect the model performance meaningfully.

The primary objective of the simulation study is to evaluate how the proportion of mediated effects affects the proposed models. Accordingly, for the first mediation model, we fixed $\boldsymbol{B}^{[x\rightarrow{y}]}$ so that the predictor account for $13\%$ variability of the growth factors of the outcome process (i.e., the medium level according to \citet[Chapter~9]{Cohen1988R}). Additionally, we considered three levels of $\boldsymbol{B}^{[x\rightarrow{m}]}$ so that the predictor explains zero, medium (i.e., $13\%$), and substantial (i.e., $26\%$) variability of the growth factors of the mediator process. We then manipulated $\boldsymbol{B}^{[m\rightarrow{y}]}$ to further adjust the proportion of mediated effects (see the setup of coefficients for each condition in Table S1 in the Online Supplementary Document). For the elements in $\boldsymbol{B}^{[m\rightarrow{y}]}$, we considered the diagonal coefficients and off-diagonal coefficients separately since the diagonal elements are related to the immediate effects\footnote{The immediate effects could be, for example, the coefficient from the first slope of the mediator process to the first slope of the outcome process or that from the knot measurement of the mediator process to the knot measurement of the outcome process.}, while the off-diagonal elements are related to the delayed effects\footnote{The delayed effects could be, for example, the coefficients from the first slope of the mediator process to the knot measurement of the outcome process.} in a mediational process. For all three levels of $\boldsymbol{B}^{[x\rightarrow{m}]}$, we considered three levels of indirectly immediate effects and two levels of indirectly delayed effects. In addition, for $\boldsymbol{B}^{[x\rightarrow{m}]}$ where the predictor explains $26\%$ variability of the growth factors of the mediator process, we considered two additional conditions to examine how varying path coefficients with fixed mediated effects affect model performance. Among all conditions, the maximum mediated effects from $x$ to the outcome growth factors are provided in the footnote of Table S1. Note that in the condition with the maximum mediated effects, the sum of the indirect effects achieved $48\%$ of the total effect, which is comparable to other methodological work such as \citet{Cheong2003mediate}.

Similarly, in the second longitudinal mediation model, we fixed $\boldsymbol{B}^{[x\rightarrow{y}]}$ so that the growth factors of the predictor process account for $13\%$ variability of the corresponding immediate growth factor and $3.25\%$ variability of each delayed growth factor of the outcome process. We considered two levels of $\boldsymbol{B}^{[x\rightarrow{m}]}$ so that the growth factors of the predictor process account for $13\%$ ($26\%$) variability of the corresponding immediate growth factor and $3.25\%$ ($6.50\%$) variability of each delayed growth factor of the mediator process (see the setup of coefficients for each condition in Table S1 in the Online Supplementary Document). For each level of $\boldsymbol{B}^{[x\rightarrow{m}]}$, we assessed two levels of non-zero coefficients for the elements in $\boldsymbol{B}^{[m\rightarrow{y}]}$. Among all conditions, the maximum mediated effects from the growth factors of the predictor process to those of the outcome processes are provided in Table S2. Note that in the condition with the maximum mediated effects, the sum of the indirect effects achieved $51\%$ of the total effect, which is comparable to other methodological work such as \citet{Cheong2003mediate}.

Another important factor in the simulation design to evaluate longitudinal models is the number of repeated measures since the proposed model is employed to analyze a longitudinal data set. Intuitively, the model should perform better with more repeated measurements. We also realized that the transition time to the second stage of the covariate (mediator) process, ideally, should occur no later than that of the mediator (outcome) process in longitudinal mediation models. We then decided to test two different levels of the number of measurements: six and ten. We selected six as the minimum number of repeated measurements to ensure the proposed model was fully identified\footnote{\citet[Chapter~4]{Bollen2005LCM} has shown that the latent growth model with linear-linear functional form can be identified with at least five waves with a knot at the midway of the study duration, though no studies provided information on identification for the model with an unknown knot.}. For the conditions with six repeated measures, we set the transition time to the second stage of the mediator and outcome process at halfway through the study duration ($\mu^{[u]}_{\gamma}=2.5$, $u=x,m,y$). We considered the other condition, ten measurement occasions, for two reasons. On the one hand, we wanted to examine whether an increasing number of repeated measures would improve model performance. More importantly, this condition allowed us to evaluate the model performance under two scenarios: (1) the knot of each longitudinal process is in the middle of the study duration (i.e., $\mu^{[u]}_{\gamma}=4.5$, $u=x,m,y$); (2) the transition time of the mediator (covariate) process occurs earlier than that of the outcome (mediator) process (i.e., $\mu^{[m]}_{\gamma}=3.5$ and $\mu^{[y]}_{\gamma}=5.5$ for the first model, while $\mu^{[x]}_{\gamma}=3.5$,  $\mu^{[m]}_{\gamma}=3.5$ and $\mu^{[y]}_{\gamma}=5.5$ for the second model). In addition, to account for individual measurement occasions, we allowed a time window with width $(-0.25, +0.25)$ around each wave, which is considered a `medium' deviation as in \citet{Coulombe2015ignoring}. Moreover, we examined three common change patterns, two levels of residual variances ($\theta^{[u]}_{\epsilon}=1$ or $2$) and set the residual correlation as $0.3$. We also examined the models at two levels of sample size, $n=200$ and $n=500$.

We evaluate the proposed longitudinal mediation models through the performance measures, including the relative bias, the empirical standard error (SE), the relative root-mean-square error (RMSE), and the empirical coverage probability for a nominal $95\%$ confidence interval of each parameter of the proposed models. Table \ref{tbl:metric} lists the definitions and estimates of these four performance measures. Specifically, the relative bias and empirical SE quantify whether the model, on average, targets the population values and whether estimates are precise, respectively. The relative RMSE integrates the bias and the precision metric into one measure. The coverage probability tells how well the interval estimate covers the corresponding population value.

\tablehere{2}

We decided to replicate the simulation study $S=1,000$ by an empirical approach following \citet{Morris2019simulation}. Specifically, we run a pilot simulation study, the standard error of bias of all parameters except the (unexplained) variance of knot measurement of each longitudinal process (i.e., $\psi_{\gamma\gamma}^{[u]}$, $u=x,m,y$) was less than $0.15$. Therefore, we need at least $900$ replications to keep the Monte Carlo standard error of the bias below $0.005$\footnote{$\text{Monte Carlo SE(Bias)}=\sqrt{Var(\hat{\theta})/S}$ \citep{Morris2019simulation}.}. Out of more conservative consideration, we decided to proceed with $S=1,000$.

\subsection*{Data Generation and Simulation Step}\label{E:step}
We carried out the following general steps for the simulation study of the proposed mediation models:
\begin{enumerate}
\item Generate the baseline predictor (or growth factors of the predictor process), growth factors of the mediator process, and those of the outcome process simultaneously with the prespecified mean vectors and variance-covariance matrices (details are provided in Appendix \ref{supp:3}) using the R package \textit{MASS} \citep{Venables2002Statistics},
\vspace{-3mm}
\item Generate a scaled and equally-spaced time structure with $J$ waves $t_{j}$ and obtain individual measurement occasions: $t_{ij}\sim U(t_{j}-0.25, t_{j}+0.25)$ by allowing a time-window with width $(-0.25, 0.25)$ around each wave,
\vspace{-3mm}
\item Calculate factor loadings for each individual of the (predictor or) mediator or outcome process from individual measurement occasions and the process-specific knot location,
\vspace{-3mm}
\item Calculate the values of the repeated measurements for each longitudinal process from corresponding growth factors, factor loadings, as well as residual variances and covariance,
\vspace{-3mm}
\item Implement the proposed models on the generated data set, estimate the parameters, and construct corresponding $95\%$ Wald CIs,
\vspace{-3mm}
\item Repeat the steps as mentioned above until having $1,000$ convergent solutions.
\end{enumerate}

\section*{Results}\label{results}
\subsection*{Model Convergence}\label{R:Preliminary}
We first examined the convergence\footnote{The convergence was defined as arriving at \textit{OpenMx} status code $0$ that indicates a successful optimization until up to $10$ runs with different sets of initial values \citep{OpenMx2016package}.} rate under each condition for each model before evaluating how the proposed longitudinal mediation models performed. The proposed model converged satisfactorily. For the first model, where we have baseline covariate and longitudinal mediator and outcome, $365$ conditions out of $396$ conditions reported a $100\%$ convergence rate. Only $31$ condition(s) reported up to five non-convergence replications. For the second model, in which we have longitudinal covariate, mediator, and outcome, 166 conditions out of a total 180 conditions reported a $100\%$ convergence rate. Only $14$ condition(s) reported up to five non-convergence replications.

\subsection*{Performance Measures}\label{R:Primary}
We present the performance metrics, including the relative bias, empirical standard error (SE), relative root-mean-square error (RMSE), and empirical coverage probability (CP) for a nominal $95\%$ confidence interval of each parameter for the proposed models. For each parameter of interest, given the size of the conditions and parameters, we first obtained its four performance measures over $1,000$ replications under each condition in the simulation design. We then summarized these metrics across conditions as the corresponding median and range\footnote{For relative biases/RMSEs, we removed the conditions with coefficients with zero population values when summarizing the related summary statistics.}. We provide the summary of the four performance metrics for the two models in Table S3-Table S5. These tables show that both models are capable of estimating parameters unbiasedly and precisely and exhibiting target confidence interval coverage in general.

Both models yielded unbiased point estimates along with small empirical standard errors generally. Specifically, for both models, the magnitude of relative biases of the growth factor means was under $0.004$, suggesting that we are able to have unbiased mean trajectories from both models. Additionally, for the first model, the magnitude of relative biases of the coefficients from the covariate to the mediator (outcome) growth factors was under $0.03$. Yet some coefficients from mediator growth factors to outcome growth factors, for example, $\beta^{[m\rightarrow{y}]}_{12}$ and $\beta^{[m\rightarrow{y}]}_{22}$ exhibit some bias (i.e., the relative biases were greater than $10\%$). These biased coefficients further led to biased estimates for the mediated effects. For the second model, the magnitude of the coefficients from covariate growth factors to mediator (outcome) growth factors except $\beta^{[x\rightarrow{m}]}_{12}$ ($\beta^{[x\rightarrow{y}]}_{12}$) was under $0.08$. Some coefficients from the mediator growth factors to the outcome growth factors exhibit some bias.

For the first model, we provide Figures \ref{fig:rBias_M1_coef} and \ref{fig:rBias_M1_med} to further examine the relative bias pattern for the coefficients and indirect effects with some bias greater than $10\%$, respectively. From Figure \ref{fig:rBias_M1_coef}, we noticed that the estimates of $\beta^{[m\rightarrow{y}]}_{12}$ and $\beta^{[m\rightarrow{y}]}_{22}$ were satisfactory in general (i.e., less than $10\%$ relative bias). Most factors we considered in the simulation design, such as the trajectory shapes and knot locations, did not affect the relative bias meaningfully. These biased estimates were generated under the conditions with a smaller sample size ($n=200$), a shorter study duration (i.e., six measurements), and a larger residual variance (i.e., $\theta^{[u]}=2$). There are two additional findings from this figure. First, the first model tended to overestimate the path coefficient from the first slope of the mediator to the second slope of the outcome, yet to underestimate the path coefficient from the second slope of the mediator to the second slope of the outcome. Second, the magnitude of relative biases was relatively small under the conditions where there was a relatively greater relationship between the mediator and the outcome. From Figure \ref{fig:rBias_M1_med}, we also noticed that the mediated effect involved $\beta^{[m\rightarrow{y}]}_{12}$ was likely to be overestimated, while that involved $\beta^{[m\rightarrow{y}]}_{22}$ was likely to be underestimated. In addition, we noticed that varying path coefficients with fixed mediated effects might slightly affect the performance of the first model.

\figurehere{3}

\figurehere{4}

For the second model, we plot Figures \ref{fig:rBias_M2_coef} and \ref{fig:rBias_M2_med} to further examine the relative bias pattern for the coefficients and indirect effects with some bias greater than $10\%$, respectively. From Figure \ref{fig:rBias_M2_coef}, we noticed that the estimates of these coefficients were satisfactory in general (i.e., less than $10\%$). Some factors we considered in the simulation design, such as the trajectory shapes, the knot locations, and the magnitude of indirect effects, did not affect the relative bias meaningfully. These biased estimates were generated under the conditions with a smaller sample size ($n=200$), a shorter study duration (i.e., six measurements), and a larger residual variance (i.e., $\theta^{[u]}=2$). Similar to Model 1, Model 2 was likely to overestimate path coefficients from a first slope to a second slope, but to underestimate the coefficients between two second slopes. From Figure \ref{fig:rBias_M2_med}, we noticed that the mediated effects involved path coefficients from a first slope to a second slope were tended to be overestimated. In contrast, those involved path coefficients between second slopes were tended to be underestimated.

\figurehere{5}

\figurehere{6}

Moreover, estimates obtained from the proposed models were precise: the magnitude of empirical standard errors of the slope- or knot-related parameters was under $0.20$, although this value of the parameters related to the knot measurement could achieve $0.57$. The relatively large empirical SEs of the parameters related to the knot measurement were due to the large scale of their population values (the population value of intercept means was around $100$). 

The proposed models are capable of estimating parameters accurately. For both models, the magnitude of relative RMSEs of growth factor means was under $0.09$, and for knot was under $0.05$. The relative RMSE magnitude of the path coefficients, indirect effects, and direct effects was relatively large. In addition, both models performed well regarding empirical coverage since the coverage probabilities of all parameters were around $0.95$. We noticed that the coverage probabilities of indirect effects could achieve $100\%$, which is greater than the nominal coverage probability ($95\%$). As stated earlier, we constructed a $95\%$ Wald confidence interval with the SE obtained by the Delta Method and generated by \textit{mxAlgebra()} automatically for each indirect effect, which should be very similar to the Sobel SE \citep{Sobel1982indirect, Sobel1986indirect, Baron1986indirect}. Accordingly, the issue of overestimated SE is in line with an earlier study \citep{Cheung2009indirect} and within our expectations to see.

\section*{Application}\label{application}
In this section, we demonstrate how to employ the proposed longitudinal mediation models to analyze longitudinal records from the ECLS-K: $2011$. This section includes two examples. We illustrate how to apply the first model to investigate how the baseline attentional focus affects the development of reading ability and then the development of mathematics ability. We then demonstrate how to implement the second model to explore how the development of reading ability affects that of mathematics and science ability. We randomly extracted $400$ students from ECLS-K: 2011 with complete records of repeated reading, mathematics, science item response theory (IRT) scaled scores, age at each wave, and baseline attentional focus\footnote{There are $n=18174$ participants in ECLS-K: 2011 in total. After removing records with missing values (i.e., rows with any of NaN/-9/-8/-7/-1), the number of entries is $n=2290$.}.

ECLS-K: 2011 is a national longitudinal study of children registered in about $900$ kindergarten programs starting from $2010-2011$ school year in the United States. In ECLS-K: 2011, participants' reading and mathematics ability were evaluated in nine waves: fall and spring of kindergarten ($2010-2011$), first ($2011-2012$) and second ($2012-2013$) grade, respectively, as well as spring of $3^{rd}$ ($2014$), $4^{th}$ ($2015$) and $5^{th}$ ($2016$) grade, respectively. Only about $30\%$ students were evaluated in the fall of $2011$ and $2012$ \citep{Le2011ECLS}. Students' science assessment started from the spring of kindergarten; accordingly, it was only examined in eight waves. We used children's age (in months) instead of their grade-in-school to obtain individual measurement occasions in analyses. We considered baseline attentional focus (ranging between $1$ and $7$) as a continuous variable, and its corresponding mean (SD) was 4.91 (1.28) in the extracted subset. Additionally, in the subsample, $47.75\%$ and $52.25\%$ of children were boys and girls. Additionally, the extracted sample was represented by $41.75\%$ White, $9.00\%$ Black, $38.00\%$ Latinx, $6.00\%$ Asian, and $5.25\%$ others. 

\subsection*{Univariate Development}
Ideally, as stated earlier, the knot of the covariate process, indicating a milestone or an event in the process, occurs first, followed by the knot of the mediator process and then that of the outcome process. Therefore, we can specify the paths from the covariate knot to the other two knots and the path from the mediator knot to the outcome knot if we observe this temporal order. Otherwise, these paths between the knots may not be reasonable since we usually do not expect a later event predict an earlier one. Therefore, in this section, we first constructed a univariate latent growth curve model with the linear-linear functional form to estimate the fixed knot for each longitudinal process and then decide on possible paths for the proposed models. The estimated knot of the development of reading, mathematics, and science ability was around $93$, $100$, and $100$ months, respectively, suggesting that we can construct longitudinal models with the paths between these knots. 

\subsection*{Baseline Attentional Focus and Longitudinal Reading and Mathematics Ability}
In this section, we built the first longitudinal mediation model to assess how the baseline attentional focus affects the development of reading ability and then the development of mathematics ability. As shown in Figure \ref{fig:traj_model1}, the estimates obtained from the first model lead to a model trajectory that sufficiently captures the smooth line of observed trajectories for each ability. Table \ref{tbl:Est1} presents the estimates of parameters of interest for the first model. Post-knot development of both abilities slowed down substantially. The transition to the slower rate occurred earlier in reading ability ($93$ months) than in mathematics ability ($99$ months). Baseline attentional focus positively affects the pre-knot development and the knot measurement of reading ability, and the knot measurement of mathematics ability. For example, the knot measurement of reading ability and mathematics ability increased $5.897$ and $2.277$ with one standardized unit increase in baseline attentional focus. Moreover, a child who developed more rapidly in reading ability tended to develop more rapidly in mathematics; a child who had higher scores at the knot of reading development tended to have higher mathematics scores at its knot. Additionally, pre-knot development of reading affects the knot measurement of mathematics development positively (estimate was $2.871$ with the p-value of $0.0453$). 

\figurehere{7}

\tablehere{3}

In terms of mediated effects, the baseline attentional focus is associated with the pre-knot development, knot measurement, and post-knot development of the mathematics process indirectly through the corresponding counterpart of the reading process. In addition, we noticed that the effects on the post-knot development of both abilities of the baseline attentional focus were trivial. 

\subsection*{Longitudinal Reading, Mathematics and Science Ability}
In this section, we built the second longitudinal mediation model to evaluate how the development of reading ability affects the development of mathematics ability and then the development of science ability. From Figure \ref{fig:traj_model2}, the estimates obtained from the second model also lead to a model-implied trajectory that sufficiently captures the smooth line of observed values for each disciplinary subject. We list the estimates of parameters of interest for the second model in Table \ref{tbl:Est2}, where we can see post-knot development of science ability slowed down slightly. The change to the slower rate occurred earliest in reading ability ($93$ months), followed by mathematics ability ($99$ months), and then science ability ($99$ months). In general, a child who developed more rapidly in reading ability tended to develop more rapidly in mathematics (science) at an early stage, and a child who had higher scores at the knot of reading development tended to have higher mathematics (science) scores at the corresponding knot. We also noticed that pre-knot development of mathematics impacted the knot measurement of science IRT scores negatively, which seems counter-intuitive. We will provide possible explanations in the Discussion section.

\tablehere{4}

In terms of mediated effects, the pre-knot development, knot measurement, and post-knot development of the reading ability positively affected the corresponding values of science ability through the counterpart of mathematics ability. Additionally, pre-knot development of reading ability influenced the knot measurement of mathematics scores and the knot measurement of science scores. The knot measurement of reading ability also impacted the post-knot development of mathematics skills and the post-knot development of science skills. In terms of total effects, all three growth factors of reading development affected the corresponding immediate and delayed growth factors of science development. For example, the total effects of the pre-knot development of reading ability on the pre-knot development, knot measurement, and post-knot development of science ability were $0.339$, $7.347$, and $-0.330$, respectively. 

\section*{Discussion}\label{discussion}
In this article, we propose two longitudinal mediation models to evaluate how a covariate, either its baseline value or its process, affects a mediator process and thus an outcome process. We employ the bilinear spline functional form to approximate the underlying change patterns of all longitudinal processes in the proposed models. This functional form allows for exploring how the change rate during the early stage or the value at a milestone (i.e., the measurement at a knot) of the covariate (mediator) process affects the change rate during the late stage of the mediator (outcome) process. We conducted extensive simulation studies to evaluate the proposed models in terms of convergence rate and performance metrics, including the relative bias, empirical standard error, relative RMSE, and coverage probability. We also illustrate the proposed models using a real-world data set from longitudinal records of reading, mathematics, and science IRT scores from Grade $K$ to Grade $5$. The results demonstrate the models' valuable capabilities of capturing the underlying change patterns of nonlinear longitudinal processes and estimating direct and indirect effects on the development of the outcome of the covariate. 

\subsection*{Practical Considerations}\label{D:practical}
In this section, we provide a set of recommendations for empirical researchers who are interested in employing the proposed models. First, we recommend building a univariate growth curve model before constructing the proposed models, as we did in the Application section. These univariate growth curve models allow us to assess whether the proposed functional form can capture the underlying change patterns of each longitudinal process. More importantly, we obtain the estimated knot of each longitudinal process, which allows for examining whether the temporal order of the covariate knot, mediator knot, and outcome knot is satisfied. If we have the temporal order as in the Application section, it is reasonable to add the paths between these knots. Otherwise, we may want to remove the paths to avoid a logical fallacy. For example, if the outcome knot occurred earlier than the mediator knot, we may not want to have the path from the mediator knot to the outcome knot in longitudinal mediation models. 

Second, we constructed full models in the simulation study to examine model performance in estimating each possible parameter within research interests. However, we suggest only using the models with all possible paths for exploratory purposes. We recommend constructing a parsimonious model for a more confirmatory analysis due to two considerations. First, it helps avoid the potential collinearity issue in a confirmatory model, which we encountered in the Application section. In the second example, we found that the direct effects on the knot measurement of science ability of the pre-knot development of reading and mathematics ability were $9.455$ and $-8.667$, respectively. Although pre-knot development of mathematics ability might negatively affect the knot measurement of science ability, a more reasonable explanation is that the pre-knot development of reading and mathematics abilities were highly correlated. Therefore, the estimate of either individual effect was invalid, although the bundle of the pre-knot development of the two abilities can still predict the knot measurement of science ability. Accordingly, the paths in a more confirmatory model, where the individual effect is of interest, need to be carefully determined by specific research questions. More importantly, a parsimonious model with fewer paths may help model identification by decreasing the likelihood of the appearance of \textit{a not positive-definite sample covariance matrix}, a common error in all models in the SEM framework. One underlying mechanism is related to the proposition regarding the positive definite property of the symmetric block matrix, as demonstrated in Appendix \ref{supp:3}).

Third, the models proposed and the analyses performs in the two examples in the Application section were pedagogical. We analyzed how the baseline attentional focus affects mathematics development through reading development by constructing the first model. Although we can maintain the temporal order, we cannot claim causality. The alternative pathway that is attentional focus affects the mathematics ability and then influences the reading ability is also plausible. Similarly, an alternative pathway for the second model could be that the development of mathematics development affects science development via the mediation of reading development. Moreover, in the proposed models, only the relationship between those between-construct growth factors is captured by unidirectional paths; that is, the development during the first phase of one ability is not considered a predictor of the development during the second phase of the same ability. Therefore, Equations \ref{eq:M1_Mvar}, \ref{eq:M1_Yvar}, \ref{eq:M2_Mvar} and \ref{eq:M2_Yvar} are helpful to obtain the covariance (correlation) between such within-construct growth factors if it is of research interest. In addition, similar to all mediation models, other confounders, such as learning approach or socioeconomic status, could also explain the relationship.

Last, researchers should be aware of the multiple statistical tests conducted when fitting a longitudinal mediation model and should consider adjusting hypothesis tests to control the Type I error rate or false discovery rate. Furthermore, decisions about the importance of insights should not only be based on p-values but also consider effect sizes, prior evidence, and alternative explanations \citep{Wasserstein2019Pvalue}. For instance, in the first example, the effects size of the baseline attentional focus on the post-knot development of reading ability was $-0.047$. Although it is statistically significant at the $0.05$ level, this finding may not be meaningful in practice due to the trivial effect size. Additionally, one should carefully interpret the effect of the pre-knot slope of the covariate to the post-knot slope of the mediator (outcome), especially under conditions with fewer measurement occasions and less precise measurement where the estimates could exhibit some bias greater than $10\%$ based on our simulation study. 

\subsection*{Methodological Considerations and Future Directions}\label{D:method}
There are several directions to consider for future studies. First, the present study allows for employing the bilinear spline functional form to approximate nonlinear change patterns. This functional form is versatile and valuable for exploring longitudinal processes where different change rates correspond to different stages. More importantly, the estimated knot, which indicates a milestone or an event of a longitudinal process, is especially useful in a mediation model because it helps avoid a logical fallacy and ensures consistency between theory and statistical methods. Although the linear-linear piecewise functional form is sufficient in most cases, such as a developmental process with earlier and later stages in psychological phenomena or a recovery process with short-term and long-term recovery periods in biomedical fields, a functional form with more linear segments may also prove useful in practice. The proposed models can also be extended accordingly. 

Second, we constructed the Wald confidence interval (CI) with the SE obtained by the Delta Method for the indirect effects and the total effects in the simulation studies and real-world data analyses. In the simulation studies, we noticed that the coverage probabilities could achieve $100\%$, greater than the nominal coverage probability ($95\%$). Another possible future direction is to construct other types of confidence intervals, such as the bootstrap CIs. Additionally, it is worth conducting simulation studies to evaluate these alternative methods for constructing CIs. 

Third, we may not want to include all possible paths to construct a parsimonious model and avoid the potential collinearity issue. Although we recommend building the reduced model driven by specific research questions, it is still worth developing hypothesis testing for comparing the full model and reduced models, especially in an exploratory stage where even empirical researchers only have vague assumptions about causal relationships. In addition, the development of longitudinal mediation models that allow for penalization and regularization, either through $L_{1}$ norm (i.e., least absolute shrinkage and selection operator, LASSO \citep{Santosa1986LASSO}), or $L_{2}$ norm (i.e., ridge adjustment \citep{Hoerl1970Ridge1, Hoerl1970Ridge2}), or both (i.e., elastic net regularization \citep{Zou2005Elastic}), to automatically select the most important paths and/or shrink the estimates for highly correlated paths is an alternative solution to the issue of collinearity and overparameterization. For a specific research question, after the most important paths and then the most important mediated effects are identified, another possible future direction is to perform power analysis with these identified paths and more levels of sample size and provide some guidelines for future studies.

\subsection*{Concluding Remarks}\label{D:conclude}
To summarize, in this article, we propose two longitudinal mediation models to explore multiple nonlinear longitudinal processes. We can evaluate how the baseline covariate or the covariate process affects the outcome process through the mediator process with the proposed models. As discussed above, the proposed methods can be further extended in practice and further examined in methodology. 

\appendix
\renewcommand{\thesection}{Appendix \Alph{section}}
\renewcommand{\thesubsection}{A.\arabic{subsection}}

\section{Expressions of Additional Parameters in the First Proposed Model}\label{supp:1}
\renewcommand{\theequation}{A.\arabic{equation}}
\setcounter{equation}{0}
For the first mediational model, we express the expected mean vector and the variance-covariance matrix of the mediator and outcome process for the $i^{th}$ individual as
\begin{equation}\nonumber
\boldsymbol{\mu}_{i}=\begin{pmatrix}
\boldsymbol{\mu}^{[m]}_{i} \\ \boldsymbol{\mu}^{[y]}_{i}
\end{pmatrix}=\begin{pmatrix}
\boldsymbol{\Lambda}_{i}^{[m]} & \boldsymbol{0} \\ \boldsymbol{0} & \boldsymbol{\Lambda}_{i}^{[y]}
\end{pmatrix}\times\begin{pmatrix}
\boldsymbol{\mu}^{[m]}_{\boldsymbol{\eta}} \\ \boldsymbol{\mu}^{[y]}_{\boldsymbol{\eta}}
\end{pmatrix}
\end{equation}
and
\begin{equation}\nonumber
\begin{aligned}
\boldsymbol{\Sigma}_{i}&=\begin{pmatrix}
\boldsymbol{\Sigma}^{[m]}_{i} & \boldsymbol{\Sigma}^{[my]}_{i} \\
& \boldsymbol{\Sigma}^{[y]}_{i}
\end{pmatrix}\\
&=\begin{pmatrix}
\boldsymbol{\Lambda}_{i}^{[m]} & \boldsymbol{0} \\ \boldsymbol{0} & \boldsymbol{\Lambda}_{i}^{[y]}
\end{pmatrix}\times\begin{pmatrix}
\text{Var}(\boldsymbol{\eta}^{[m]}) & \boldsymbol{0} \\
& \text{Var}(\boldsymbol{\eta}^{[y]})
\end{pmatrix}\times\begin{pmatrix}
\boldsymbol{\Lambda}_{i}^{[m]} & \boldsymbol{0} \\ \boldsymbol{0} & \boldsymbol{\Lambda}_{i}^{[y]}
\end{pmatrix}^{T}+\begin{pmatrix}
\theta^{[m]}_{\epsilon}\boldsymbol{I} & \theta^{[my]}_{\epsilon}\boldsymbol{I} \\
& \theta^{[y]}_{\epsilon}\boldsymbol{I}
\end{pmatrix},
\end{aligned}
\end{equation}
where $\boldsymbol{\mu}^{[m]}_{\boldsymbol{\eta}}$ and $\text{Var}(\boldsymbol{\eta}^{[m]})$ are the conditional mean vector and variance-covariance matrix of the mediator's growth factors given the covariate, which can be expressed as
\begin{align}
&\boldsymbol{\mu}^{[m]}_{\boldsymbol{\eta}}=\boldsymbol{\alpha}^{[m]}+\boldsymbol{B}^{[x\rightarrow{m}]}\mu_{x}\label{eq:M1_Mmean},\\
&\text{Var}(\boldsymbol{\eta}^{[m]})=\boldsymbol{\Psi}^{[m]}_{\boldsymbol{\eta}}+\boldsymbol{B}^{[x\rightarrow{m}]}\Phi_{x}\boldsymbol{B}^{[x\rightarrow{m}]T}\label{eq:M1_Mvar},
\end{align}
in which $\mu_{x}$ and $\Phi_{x}$ are the mean value and variance of the covariate. Additionally, $\boldsymbol{\mu}^{[y]}_{\boldsymbol{\eta}}$ and $\text{Var}(\boldsymbol{\eta}^{[y]})$ are the conditional mean vector and variance-covariance matrix of the growth factors of the outcome progress given the covariate and mediator progress
\begin{align}
&\boldsymbol{\mu}^{[y]}_{\boldsymbol{\eta}}=\boldsymbol{\alpha}^{[y]}+\boldsymbol{B}^{[x\rightarrow{y}]}\mu_{x}+\boldsymbol{B}^{[m\rightarrow{y}]}\boldsymbol{\mu}^{[m]}_{\boldsymbol{\eta}}\label{eq:M1_Ymean}\\
&\text{Var}(\boldsymbol{\eta}^{[y]})=\boldsymbol{\Psi}^{[y]}_{\boldsymbol{\eta}}+\boldsymbol{B}^{[x\rightarrow{y}]}\Phi_{x}\boldsymbol{B}^{[x\rightarrow{y}]T}+\boldsymbol{B}^{[m\rightarrow{y}]}\text{Var}(\boldsymbol{\eta}^{[m]})\boldsymbol{B}^{[m\rightarrow{y}]T}\label{eq:M1_Yvar}.
\end{align}

\newpage
\section{Expressions of Additional Parameters in the Second Proposed Model}\label{supp:2}
\renewcommand{\theequation}{B.\arabic{equation}}
\setcounter{equation}{0}
For the second mediational model, we express the expected mean vector and the variance-covariance matrix of the three longitudinal processes for individual $i$ as
\begin{equation}\nonumber
\boldsymbol{\mu}_{i}=\begin{pmatrix}
\boldsymbol{\mu}^{[x]}_{i} \\ \boldsymbol{\mu}^{[m]}_{i} \\ \boldsymbol{\mu}^{[y]}_{i}
\end{pmatrix}=\begin{pmatrix}
\boldsymbol{\Lambda}_{i}^{[x]} & \boldsymbol{0} & \boldsymbol{0} \\
\boldsymbol{0} & \boldsymbol{\Lambda}_{i}^{[m]} & \boldsymbol{0} \\ 
\boldsymbol{0} & \boldsymbol{0} & \boldsymbol{\Lambda}_{i}^{[y]}
\end{pmatrix}\times\begin{pmatrix}
\boldsymbol{\mu}^{[x]}_{\boldsymbol{\eta}} \\ \boldsymbol{\mu}^{[m]}_{\boldsymbol{\eta}} \\ \boldsymbol{\mu}^{[y]}_{\boldsymbol{\eta}}
\end{pmatrix}
\end{equation}
and
\begin{equation}\nonumber
\begin{footnotesize}
\begin{aligned}
\boldsymbol{\Sigma}_{i}&=\begin{pmatrix}
\boldsymbol{\Sigma}^{[x]}_{i} & \boldsymbol{\Sigma}^{[xm]}_{i} & \boldsymbol{\Sigma}^{[xy]}_{i} \\
& \boldsymbol{\Sigma}^{[m]}_{i} & \boldsymbol{\Sigma}^{[my]}_{i} \\
& & \boldsymbol{\Sigma}^{[y]}_{i}
\end{pmatrix}\\
&=\begin{pmatrix}
\boldsymbol{\Lambda}_{i}^{[x]} & \boldsymbol{0} & \boldsymbol{0} \\
\boldsymbol{0} & \boldsymbol{\Lambda}_{i}^{[m]} & \boldsymbol{0} \\ 
\boldsymbol{0} & \boldsymbol{0} & \boldsymbol{\Lambda}_{i}^{[y]}
\end{pmatrix}\times\begin{pmatrix}
\boldsymbol{\Psi}^{[x]}_{\boldsymbol{\eta}} & \boldsymbol{0} & \boldsymbol{0} \\
& \text{Var}(\boldsymbol{\eta}^{[m]}) & \boldsymbol{0} \\
& & \text{Var}(\boldsymbol{\eta}^{[y]})
\end{pmatrix}\times\begin{pmatrix}
\boldsymbol{\Lambda}_{i}^{[x]} & \boldsymbol{0} & \boldsymbol{0} \\
\boldsymbol{0} & \boldsymbol{\Lambda}_{i}^{[m]} & \boldsymbol{0} \\ 
\boldsymbol{0} & \boldsymbol{0} & \boldsymbol{\Lambda}_{i}^{[y]}
\end{pmatrix}^{T}+\begin{pmatrix}
\theta^{[x]}_{\epsilon}\boldsymbol{I} & \theta^{[xm]}_{\epsilon}\boldsymbol{I} & \theta^{[xy]}_{\epsilon}\boldsymbol{I} \\
& \theta^{[m]}_{\epsilon}\boldsymbol{I} & \theta^{[my]}_{\epsilon}\boldsymbol{I} \\
& & \theta^{[y]}_{\epsilon}\boldsymbol{I}
\end{pmatrix},
\end{aligned}
\end{footnotesize}
\end{equation}
where $\boldsymbol{\mu}^{[x]}_{\boldsymbol{\eta}}$ and $\boldsymbol{\Psi}^{[x]}_{\boldsymbol{\eta}}$ are the mean vector and variance-covariance matrix of the growth factors of the covariate process, $\boldsymbol{\mu}^{[m]}_{\boldsymbol{\eta}}$ and $\text{Var}(\boldsymbol{\eta}^{[m]})$ are the conditional mean vector and variance-covariance matrix of the growth factors of the mediator given those of the covariate, which can be expressed as
\begin{align}
&\boldsymbol{\mu}^{[m]}_{\boldsymbol{\eta}}=\boldsymbol{\alpha}^{[m]}+\boldsymbol{B}^{[x\rightarrow{m}]}\boldsymbol{\mu}^{[x]}_{\boldsymbol{\eta}},\label{eq:M2_Mmean}\\
&\text{Var}(\boldsymbol{\eta}^{[m]})=\boldsymbol{\Psi}^{[m]}_{\boldsymbol{\eta}}+\boldsymbol{B}^{[x\rightarrow{m}]}\boldsymbol{\Psi}^{[x]}_{\boldsymbol{\eta}}\boldsymbol{B}^{[x\rightarrow{m}]T},\label{eq:M2_Mvar}
\end{align}
Additionally, $\boldsymbol{\mu}^{[y]}_{\boldsymbol{\eta}}$ and $\text{Var}(\boldsymbol{\eta}^{[y]})$ are the conditional mean vector and variance-covariance matrix of the growth factors of the outcome progress given those of the covariate and mediator progress, which can be written as follows
\begin{align}
&\boldsymbol{\mu}^{[y]}_{\boldsymbol{\eta}}=\boldsymbol{\alpha}^{[y]}+\boldsymbol{B}^{[x\rightarrow{y}]}\boldsymbol{\mu}^{[x]}_{\boldsymbol{\eta}}+\boldsymbol{B}^{[m\rightarrow{y}]}\boldsymbol{\mu}^{[m]}_{\boldsymbol{\eta}},\label{eq:M2_Ymean}\\
&\text{Var}(\boldsymbol{\eta}^{[y]})=\boldsymbol{\Psi}^{[y]}_{\boldsymbol{\eta}}+\boldsymbol{B}^{[x\rightarrow{y}]}\boldsymbol{\Psi}^{[x]}\boldsymbol{B}^{[x\rightarrow{y}]T}+\boldsymbol{B}^{[m\rightarrow{y}]}\text{Var}(\boldsymbol{\eta}^{[m]})\boldsymbol{B}^{[m\rightarrow{y}]T}.\label{eq:M2_Yvar}
\end{align}

\section{Data Generation Approach and Its Implication to Model Identification}\label{supp:3}
\renewcommand{\theequation}{C.\arabic{equation}}
\setcounter{equation}{0}
There are two approaches to generate a data set for a regression model. Take the simple linear regression as an example. The first approach is to generate a predictor $\boldsymbol{x}$ and then generate a conditional distribution of $\boldsymbol{y}$ on $\boldsymbol{x}$ with pre-specified coefficients. Alternatively, we can generate a joint distribution of $\boldsymbol{x}$ and $\boldsymbol{y}$ simultaneously. In the present article, we generated a joint distribution for the baseline predictor (or growth factor of the predictor process), growth factors of the mediator process, and the growth factors of the outcome process. Therefore, we simultaneously generate the mean vector and variance-covariance matrix for all parameters involved in the mediation process for each proposed model. One challenge of the approach is that we need to express the mean vector and variance-covariance matrix. This appendix provides the expressions of the mean vector and variance-covariance matrix for both models.

The parameters involved in the first mediation model include the mean and variance of the baseline predictor, the mean vector and variance-covariance matrix of the growth factors of the mediator process, and the mean vector and variance-covariance matrix of the growth factors of the outcome process. The mean vector of these parameters can be expressed as
\begin{equation}\nonumber
\boldsymbol{\mu}_{\text{Model}_{1}}=\begin{pmatrix}
\boldsymbol{\mu}^{[y]}_{\boldsymbol{\eta}} & \boldsymbol{\mu}^{[m]}_{\boldsymbol{\eta}} & \mu_{x}
\end{pmatrix}^{T},
\end{equation}
where $\boldsymbol{\mu}^{[y]}_{\boldsymbol{\eta}}$, $\boldsymbol{\mu}^{[m]}_{\boldsymbol{\eta}}$, and $\mu_{x}$ have the same definitions as those in Equations \ref{eq:M1_Mmean} and \ref{eq:M1_Ymean}. The variance-covariance matrix of these parameters is 
\begin{equation}\label{Model1_var}
\boldsymbol{\Sigma}_{\text{Model}_{1}}=\begin{pmatrix}
\text{Var}(\boldsymbol{\eta}^{[y]}) & \boldsymbol{B}^{[m\rightarrow y]}\text{Var}(\boldsymbol{\eta}^{[m]})+\boldsymbol{B}^{[x\rightarrow y]}\Phi_{x}\boldsymbol{B}^{[x\rightarrow m]T} & \boldsymbol{B}^{[m\rightarrow y]}\boldsymbol{B}^{[x\rightarrow m]}\Phi_{x}+\boldsymbol{B}^{[x\rightarrow y]}\Phi_{x} \\
 & \text{Var}(\boldsymbol{\eta}^{[m]}) & \boldsymbol{B}^{[x\rightarrow m]}\Phi_{x} \\
 & & \Phi_{x}
\end{pmatrix},
\end{equation}
where $\text{Var}(\boldsymbol{\eta}^{[y]})$, $\text{Var}(\boldsymbol{\eta}^{[m]})$, and $\Phi_{x}$ have the same definitions as those in Equations \ref{eq:M1_Mvar} and \ref{eq:M1_Yvar}. Therefore, the joint distribution of all parameters involved in the first mediation longitudinal model follows a multivariate normal distribution $N_{7}(\boldsymbol{\mu}_{\text{Model}_{1}}, \boldsymbol{\Sigma}_{\text{Model}_{1}})$.

The parameters involved in the second mediation model include the mean vector and variance-covariance matrix of the growth factors of the predictor process, the mean vector and variance-covariance matrix of the growth factors of the mediator process, and the mean vector and variance-covariance matrix of the growth factors of the outcome process. The mean vector of these parameters can be expressed as
\begin{equation}\nonumber
\boldsymbol{\mu}_{\text{Model}_{2}}=\begin{pmatrix}
\boldsymbol{\mu}^{[y]}_{\boldsymbol{\eta}} & \boldsymbol{\mu}^{[m]}_{\boldsymbol{\eta}} & \boldsymbol{\mu}^{[x]}_{\boldsymbol{\eta}}
\end{pmatrix}^{T},
\end{equation}
where $\boldsymbol{\mu}^{[y]}_{\boldsymbol{\eta}}$, $\boldsymbol{\mu}^{[m]}_{\boldsymbol{\eta}}$, and $\boldsymbol{\mu}^{[x]}_{\boldsymbol{\eta}}$ have the same definitions as those in Equations \ref{eq:M2_Mmean} and \ref{eq:M2_Ymean}. The variance-covariance matrix of these parameters is 
\begin{equation}\label{Model2_var}
\boldsymbol{\Sigma}_{\text{Model}_{2}}=\begin{pmatrix}
\text{Var}(\boldsymbol{\eta}^{[y]}) & \boldsymbol{B}^{[m\rightarrow y]}\text{Var}(\boldsymbol{\eta}^{[m]})+\boldsymbol{B}^{[x\rightarrow y]}\boldsymbol{\Psi}^{[x]}_{\boldsymbol{\eta}}\boldsymbol{B}^{[x\rightarrow m]T} & \boldsymbol{B}^{[m\rightarrow y]}\boldsymbol{B}^{[x\rightarrow m]}\boldsymbol{\Psi}^{[x]}_{\boldsymbol{\eta}}+\boldsymbol{B}^{[x\rightarrow y]}\boldsymbol{\Psi}^{[x]}_{\boldsymbol{\eta}} \\
 & \text{Var}(\boldsymbol{\eta}^{[m]}) & \boldsymbol{B}^{[x\rightarrow m]}\boldsymbol{\Psi}^{[x]}_{\boldsymbol{\eta}} \\
 & & \boldsymbol{\Psi}^{[x]}_{\boldsymbol{\eta}}
\end{pmatrix},
\end{equation}
where $\text{Var}(\boldsymbol{\eta}^{[y]})$, $\text{Var}(\boldsymbol{\eta}^{[m]})$, and $\boldsymbol{\Psi}^{[x]}_{\boldsymbol{\eta}}$ have the same definitions as those in Equations \ref{eq:M2_Mvar} and \ref{eq:M2_Yvar}. Therefore, the joint distribution of all parameters involved in the second longitudinal mediation model follows a multivariate normal distribution $N_{9}(\boldsymbol{\mu}_{\text{Model}_{2}}, \boldsymbol{\Sigma}_{\text{Model}_{2}})$.

One condition of generating the joint distribution of each model is that the corresponding variance-covariance matrix should be a positive definite matrix. According to the proposition regarding the positive definite property of the symmetric block matrix, conceptually, the variance-covariance matrices defined in Equations \ref{Model1_var} and \ref{Model2_var} may not be positive definite if we assign large values to the path coefficients $\boldsymbol{B}^{[x\rightarrow m]}$, $\boldsymbol{B}^{[x\rightarrow y]}$, or $\boldsymbol{B}^{[m\rightarrow y]}$. This implies that the proposed models may suffer the non-convergence issue or fail to converge to global optima if these coefficients are large.

\renewcommand\thetable{\arabic{table}}
\setcounter{table}{0}

\begin{table}
\centering
\begin{threeparttable}
\caption{Possible Indirect Effects and Total Effects on Outcome Process of Covariate}
\begin{tabular}{p{5cm}rr}
\hline
\hline
\multicolumn{3}{c}{\textbf{Baseline Predictor$\rightarrow$Longitudinal Mediator$\rightarrow$Longitudinal Outcome}} \\
\hline
\multicolumn{3}{c}{\textbf{Indirect Effects}} \\
\hline
\textbf{Putative Mediator} & \textbf{Indirect Effects} & \textbf{Estimates} \\
\hline
\multirow{3}{*}{$\boldsymbol{\eta^{[m]}_{1}}$} & $x\rightarrow{\eta^{[m]}_{1}}\rightarrow{\eta^{[y]}_{1}}$ & $\beta^{[x\rightarrow{m}]}_{1}\times\beta^{[m\rightarrow{y}]}_{11}$ \\
& $x\rightarrow{\eta^{[m]}_{1}}\rightarrow{\eta^{[y]}_{\gamma}}$ & $\beta^{[x\rightarrow{m}]}_{1}\times\beta^{[m\rightarrow{y}]}_{1\gamma}$ \\
& $x\rightarrow{\eta^{[m]}_{1}}\rightarrow{\eta^{[y]}_{2}}$ & $\beta^{[x\rightarrow{m}]}_{1}\times\beta^{[m\rightarrow{y}]}_{12}$ \\
\hline
\multirow{2}{*}{$\boldsymbol{\eta^{[m]}_{\gamma}}$} & $x\rightarrow{\eta^{[m]}_{\gamma}}\rightarrow{\eta^{[y]}_{\gamma}}$ & $\beta^{[x\rightarrow{m}]}_{\gamma}\times\beta^{[m\rightarrow{y}]}_{\gamma\gamma}$\\
& $x\rightarrow{\eta^{[m]}_{\gamma}}\rightarrow{\eta^{[y]}_{2}}$ & $\beta^{[x\rightarrow{m}]}_{\gamma}\times\beta^{[m\rightarrow{y}]}_{\gamma2}$\\
\hline
\textbf{$\boldsymbol{\eta^{[m]}_{2}}$} & $x\rightarrow{\eta^{[m]}_{2}}\rightarrow{\eta^{[y]}_{2}}$ & $\beta^{[x\rightarrow{m}]}_{2}\times\beta^{[m\rightarrow{y}]}_{22}$\\
\hline
\multicolumn{3}{c}{\textbf{Total Effects}} \\
\hline
\textbf{Total Effects} & \multicolumn{2}{c}{\textbf{Estimates}} \\
\hline
$\boldsymbol{x\rightarrow{\eta^{[y]}_{1}}}$ & \multicolumn{2}{r}{$\beta^{[x\rightarrow{y}]}_{1}+\beta^{[x\rightarrow{m}]}_{1}\times\beta^{[m\rightarrow{y}]}_{11}$} \\
\hline
$\boldsymbol{x\rightarrow{\eta^{[y]}_{\gamma}}}$ & \multicolumn{2}{r}{$\beta^{[x\rightarrow{y}]}_{\gamma}+\beta^{[x\rightarrow{m}]}_{1}\times\beta^{[m\rightarrow{y}]}_{1\gamma}+\beta^{[x\rightarrow{m}]}_{\gamma}\times\beta^{[m\rightarrow{y}]}_{\gamma\gamma}$} \\
\hline
$\boldsymbol{x\rightarrow{\eta^{[y]}_{2}}}$ & \multicolumn{2}{r}{$\beta^{[x\rightarrow{y}]}_{2}+\beta^{[x\rightarrow{m}]}_{1}\times\beta^{[m\rightarrow{y}]}_{12}+\beta^{[x\rightarrow{m}]}_{\gamma}\times\beta^{[m\rightarrow{y}]}_{\gamma2}+\beta^{[x\rightarrow{m}]}_{2}\times\beta^{[m\rightarrow{y}]}_{22}$} \\
\hline
\hline
\multicolumn{3}{c}{\textbf{Longitudinal Predictor$\rightarrow$Longitudinal Mediator$\rightarrow$Longitudinal Outcome}} \\
\hline
\multicolumn{3}{c}{\textbf{Indirect Effects}} \\
\hline
\textbf{Putative Mediator} & \textbf{Indirect Effects} & \textbf{Estimates} \\
\hline
\multirow{3}{*}{$\boldsymbol{\eta^{[m]}_{1}}$} & $\eta^{[x]}_{1}\rightarrow{\eta^{[m]}_{1}}\rightarrow{\eta^{[y]}_{1}}$ & $\beta^{[x\rightarrow{m}]}_{11}\times\beta^{[m\rightarrow{y}]}_{11}$ \\
& $\eta^{[x]}_{1}\rightarrow{\eta^{[m]}_{1}}\rightarrow{\eta^{[y]}_{\gamma}}$ & $\beta^{[x\rightarrow{m}]}_{11}\times\beta^{[m\rightarrow{y}]}_{1\gamma}$ \\
& $\eta^{[x]}_{1}\rightarrow{\eta^{[m]}_{1}}\rightarrow{\eta^{[y]}_{2}}$ & $\beta^{[x\rightarrow{m}]}_{11}\times\beta^{[m\rightarrow{y}]}_{12}$ \\
\hline
\multirow{4}{*}{$\boldsymbol{\eta^{[m]}_{\gamma}}$} & $\eta^{[x]}_{1}\rightarrow{\eta^{[m]}_{\gamma}}\rightarrow{\eta^{[y]}_{\gamma}}$ & $\beta^{[x\rightarrow{m}]}_{1\gamma}\times\beta^{[m\rightarrow{y}]}_{\gamma\gamma}$ \\
& $\eta^{[x]}_{1}\rightarrow{\eta^{[m]}_{\gamma}}\rightarrow{\eta^{[y]}_{2}}$ & $\beta^{[x\rightarrow{m}]}_{1\gamma}\times\beta^{[m\rightarrow{y}]}_{\gamma2}$ \\
& $\eta^{[x]}_{\gamma}\rightarrow{\eta^{[m]}_{\gamma}}\rightarrow{\eta^{[y]}_{\gamma}}$ & $\beta^{[x\rightarrow{m}]}_{\gamma\gamma}\times\beta^{[m\rightarrow{y}]}_{\gamma\gamma}$\\
& $\eta^{[x]}_{\gamma}\rightarrow{\eta^{[m]}_{\gamma}}\rightarrow{\eta^{[y]}_{2}}$ & $\beta^{[x\rightarrow{m}]}_{\gamma\gamma}\times\beta^{[m\rightarrow{y}]}_{\gamma2}$\\
\hline
\multirow{3}{*}{$\boldsymbol{\eta^{[m]}_{2}}$} & $\eta^{[x]}_{1}\rightarrow{\eta^{[m]}_{2}}\rightarrow{\eta^{[y]}_{2}}$ & $\beta^{[x\rightarrow{m}]}_{12}\times\beta^{[m\rightarrow{y}]}_{22}$\\
& $\eta^{[x]}_{\gamma}\rightarrow{\eta^{[m]}_{2}}\rightarrow{\eta^{[y]}_{2}}$ & $\beta^{[x\rightarrow{m}]}_{\gamma2}\times\beta^{[m\rightarrow{y}]}_{22}$\\
& $\eta^{[x]}_{2}\rightarrow{\eta^{[m]}_{2}}\rightarrow{\eta^{[y]}_{2}}$ & $\beta^{[x\rightarrow{m}]}_{22}\times\beta^{[m\rightarrow{y}]}_{22}$\\
\hline
\multicolumn{3}{c}{\textbf{Total Effects}} \\
\hline
\textbf{Total Effects} & \multicolumn{2}{c}{\textbf{Estimates}} \\
\hline
$\boldsymbol{\eta^{[x]}_{1}\rightarrow{\eta^{[y]}_{1}}}$ & \multicolumn{2}{r}{$\beta^{[x\rightarrow{y}]}_{11}+\beta^{[x\rightarrow{m}]}_{11}\times\beta^{[m\rightarrow{y}]}_{11}$} \\
\hline
$\boldsymbol{\eta^{[x]}_{1}\rightarrow{\eta^{[y]}_{\gamma}}}$ & \multicolumn{2}{r}{$\beta^{[x\rightarrow{y}]}_{1\gamma}+\beta^{[x\rightarrow{m}]}_{11}\times\beta^{[m\rightarrow{y}]}_{1\gamma}+\beta^{[x\rightarrow{m}]}_{1\gamma}\times\beta^{[m\rightarrow{y}]}_{\gamma\gamma}$} \\
\hline
$\boldsymbol{\eta^{[x]}_{1}\rightarrow{\eta^{[y]}_{2}}}$ & \multicolumn{2}{r}{$\beta^{[x\rightarrow{y}]}_{12}+\beta^{[x\rightarrow{m}]}_{11}\times\beta^{[m\rightarrow{y}]}_{12}+\beta^{[x\rightarrow{m}]}_{1\gamma}\times\beta^{[m\rightarrow{y}]}_{\gamma2}+\beta^{[x\rightarrow{m}]}_{12}\times\beta^{[m\rightarrow{y}]}_{22}$} \\
\hline
$\boldsymbol{\eta^{[x]}_{\gamma}\rightarrow{\eta^{[y]}_{\gamma}}}$ & \multicolumn{2}{r}{$\beta^{[x\rightarrow{y}]}_{\gamma\gamma}+\beta^{[x\rightarrow{m}]}_{\gamma\gamma}\times\beta^{[m\rightarrow{y}]}_{\gamma\gamma}$}\\
\hline
$\boldsymbol{\eta^{[x]}_{\gamma}\rightarrow{\eta^{[y]}_{2}}}$ & \multicolumn{2}{r}{$\beta^{[x\rightarrow{y}]}_{\gamma2}+\beta^{[x\rightarrow{m}]}_{\gamma\gamma}\times\beta^{[m\rightarrow{y}]}_{\gamma2}+\beta^{[x\rightarrow{m}]}_{\gamma2}\times\beta^{[m\rightarrow{y}]}_{22}$} \\
\hline
$\boldsymbol{\eta^{[x]}_{2}\rightarrow{\eta^{[y]}_{2}}}$ & \multicolumn{2}{r}{$\beta^{[x\rightarrow{y}]}_{22}+\beta^{[x\rightarrow{m}]}_{22}\times\beta^{[m\rightarrow{y}]}_{22}$} \\
\hline
\hline
\end{tabular}
\label{tbl:path}
\end{threeparttable}
\end{table}
\begin{table}
\centering
\begin{threeparttable}
\caption{Performance Measures for Evaluating an Estimate ($\hat{\theta}$) of Parameter ($\theta$)}
\begin{tabular}{p{4cm}p{4.5cm}p{6cm}}
\hline
\hline
\textbf{Criteria} & \textbf{Definition} & \textbf{Estimate} \\
\hline
Relative Bias & $E_{\hat{\theta}}(\hat{\theta}-\theta)/\theta$ & $\sum_{s=1}^{S}(\hat{\theta}_{s}\tnote{a}-\theta)/\theta S\tnote{b}$ \\
Empirical SE & $\sqrt{Var(\hat{\theta})}$ & $\sqrt{\sum_{s=1}^{S}(\hat{\theta}_{s}-\bar{\theta}\tnote{c})^{2}/(S-1)}$ \\
Relative RMSE & $\sqrt{E_{\hat{\theta}}(\hat{\theta}-\theta)^{2}}/\theta$ & $\sqrt{\sum_{s=1}^{S}(\hat{\theta}_{s}-\theta)^{2}/S}/\theta$ \\
Coverage Probability & $Pr(\hat{\theta}_{\text{lower}}\le\theta\le\hat{\theta}_{\text{upper}})$ & $\sum_{s=1}^{S}I(\hat{\theta}_{\text{lower},s}\le\theta\le\hat{\theta}_{\text{upper},s})\tnote{d}/S$\\
\hline
\hline
\end{tabular}
\label{tbl:metric}
\begin{tablenotes}
\small
\item[a] {$\hat{\theta}_{s}$: the estimate of $\theta$ from the $s^{th}$ replication}\\ 
\item[b] {$S$: the number of replications and set as $1,000$ in our simulation study}\\
\item[c] {$\bar{\theta}$: the mean of $\hat{\theta}_{s}$'s across replications}\\
\item[d] {$I()$: an indicator function}
\end{tablenotes}
\end{threeparttable}
\end{table}

\begin{table}
\centering
\begin{threeparttable}
\setlength{\tabcolsep}{4pt}
\renewcommand{\arraystretch}{0.6}
\caption{Estimates of Longitudinal Mediation Model for Reading and Mathematics Ability}
\begin{tabular}{p{3cm}R{3cm}R{3cm}R{3cm}R{3cm}}
\hline
\hline
& \multicolumn{4}{c}{\textbf{Growth Factor Means}} \\
\hline
\textbf{Para.} & \multicolumn{2}{c}{\textbf{Reading Ability}} & \multicolumn{2}{c}{\textbf{Mathematics Ability}} \\
\hline
& \textbf{Est. (SE)} & \textbf{P value} & \textbf{Est. (SE)} & \textbf{P value} \\
\hline
$\mu_{\eta_{1}}$ & $2.009$ ($0.028$) & $<0.0001$ & $1.761$ ($0.020$) & $<0.0001$ \\
$\mu_{\eta_{\gamma}}$ & $108.955$ ($1.011$) & $<0.0001$ & $94.651$ ($1.024$) & $<0.0001$ \\
$\mu_{\eta_{2}}$ & $0.702$ ($0.016$) & $<0.0001$ & $0.779$ ($0.018$) & $<0.0001$ \\
$\gamma$ & $93.424$ ($0.317$) & $<0.0001$ & $99.127$ ($0.413$) & $<0.0001$ \\
\hline
\hline
& \multicolumn{4}{c}{\textbf{Growth Factor (Unexplained) Variances}} \\
\hline
\textbf{Para.} & \multicolumn{2}{c}{\textbf{Reading Ability}} & \multicolumn{2}{c}{\textbf{Mathematics Ability}} \\
\hline
& \textbf{Est. (SE)} & \textbf{P value} & \textbf{Est. (SE)} & \textbf{P value} \\
\hline
$\psi_{11}$ & $0.183$ ($0.020$) & $<0.0001$ & $0.048$ ($0.007$) & $<0.0001$ \\
$\psi_{\gamma\gamma}$ & $278.52$ ($20.765$)) & $<0.0001$ & $83.196$ ($7.403$) & $<0.0001$ \\
$\psi_{22}$ & $0.042$ ($0.006$) & $<0.0001$ & $0.029$ ($0.006$) & $<0.0001$ \\
\hline
\hline
& \multicolumn{4}{c}{\textbf{Direct Effects}} \\
\hline
\textbf{Para.} & \multicolumn{2}{c}{\textbf{Covariate to Reading Ability}} & \multicolumn{2}{c}{\textbf{Covariate to Mathematics Ability}} \\
\hline
& \textbf{Est. (SE)} & \textbf{P value} & \textbf{Est. (SE)} & \textbf{P value} \\
\hline
$x\rightarrow{\eta^{[u]}_{1}}$ & $0.127$ ($0.026$) & $<0.0001$ & $0.024$ ($0.016$) & $0.1336$ \\
$x\rightarrow{\eta^{[u]}_{\gamma}}$ & $5.897$ ($0.857$) & $<0.0001$ & $2.277$ ($0.538$) & $<0.0001$ \\
$x\rightarrow{\eta^{[u]}_{2}}$ & $-0.047$ ($0.014$) & $0.0008$ & $-0.017$ ($0.015$) & $0.2571$ \\
\hline
\textbf{Para.} & \multicolumn{4}{c}{\textbf{Growth Factors of Reading Ability to Those of Mathematics Ability}} \\
\hline
& \textbf{---} & \textbf{---} & \textbf{Est. (SE)} & \textbf{P value} \\
\hline
$\eta^{[m]}_{1}\rightarrow{\eta^{[y]}_{1}}$ & --- & --- & $0.445$ ($0.042$) & $<0.0001$ \\
$\eta^{[m]}_{1}\rightarrow{\eta^{[y]}_{\gamma}}$ & --- & --- & $2.871$ ($1.434$) & $0.0453$ \\
$\eta^{[m]}_{1}\rightarrow{\eta^{[y]}_{2}}$ & --- & --- & $0.015$ ($0.062$) & $0.8088$ \\
$\eta^{[m]}_{\gamma}\rightarrow{\eta^{[y]}_{\gamma}}$ & --- & --- & $0.688$ ($0.026$) & $<0.0001$ \\
$\eta^{[m]}_{\gamma}\rightarrow{\eta^{[y]}_{2}}$ & --- & --- & $0.003$ ($0.002$) & $0.1336$ \\
$\eta^{[m]}_{2}\rightarrow{\eta^{[y]}_{2}}$ & --- & --- & $0.676$ ($0.165$) & $<0.0001$ \\
\hline
\hline
\textbf{Para.} & \multicolumn{4}{c}{\textbf{Indirect Effects}} \\
\hline
& \textbf{---} & \textbf{---} & \textbf{Est. (SE)} & \textbf{P value} \\
\hline
$x\rightarrow{\eta^{[m]}_{1}}\rightarrow{\eta^{[y]}_{1}}$ & --- & --- & $0.057$ ($0.013$) & $<0.0001$ \\
$x\rightarrow{\eta^{[m]}_{1}}\rightarrow{\eta^{[y]}_{\gamma}}$ & --- & --- & $0.366$ ($0.197$) & $0.0632$ \\
$x\rightarrow{\eta^{[m]}_{1}}\rightarrow{\eta^{[y]}_{2}}$ & --- & --- & $0.002$ ($0.008$) & $0.8026$ \\
$x\rightarrow{\eta^{[m]}_{\gamma}}\rightarrow{\eta^{[y]}_{\gamma}}$ & --- & --- & $4.055$ ($0.606$) & $<0.0001$ \\
$x\rightarrow{\eta^{[m]}_{\gamma}}\rightarrow{\eta^{[y]}_{2}}$ & --- & --- & $0.016$ ($0.012$) & $0.1824$ \\
$x\rightarrow{\eta^{[m]}_{2}}\rightarrow{\eta^{[y]}_{2}}$ & --- & --- & $-0.031$ ($0.012$) & $0.0098$ \\
\hline
\hline
\textbf{Para.} & \multicolumn{4}{c}{\textbf{Total Effects}} \\
\hline
& \textbf{---} & \textbf{---} & \textbf{Est. (SE)} & \textbf{P value} \\
\hline
$x\rightarrow{\eta^{[y]}_{1}}$ & --- & --- & $0.081$ ($0.018$) & $<0.0001$ \\
$x\rightarrow{\eta^{[y]}_{\gamma}}$ & --- & --- & $6.698$ ($0.786$) & $<0.0001$ \\
$x\rightarrow{\eta^{[y]}_{2}}$ & --- & --- & $-0.031$ ($0.015$) & $0.0388$ \\
\hline
\hline
\end{tabular}
\label{tbl:Est1}
\end{threeparttable}
\end{table}

\begin{table}
\centering
\footnotesize
\begin{threeparttable}
\setlength{\tabcolsep}{4pt}
\renewcommand{\arraystretch}{0.6}
\caption{Estimates of Longitudinal Mediation Model for Reading, Mathematics and Science Ability}
\begin{tabular}{p{2.8cm}R{2.6cm}R{1.8cm}R{2.6cm}R{1.8cm}R{2.6cm}R{1.8cm}}
\hline
\hline
& \multicolumn{6}{c}{\textbf{Growth Factor Means}} \\
\hline
\textbf{Para.} & \multicolumn{2}{c}{\textbf{Reading Ability}} & \multicolumn{2}{c}{\textbf{Mathematics Ability}} & \multicolumn{2}{c}{\textbf{Science Ability}} \\
\hline
& \textbf{Est. (SE)} & \textbf{P value} & \textbf{Est. (SE)} & \textbf{P value} \\
\hline
$\mu_{\eta_{1}}$ & $2.006$ ($0.028$) & $<0.0001$ & $1.762$ ($0.020$) & $<0.0001$ & $0.814$ ($0.014$) & $<0.0001$ \\
$\mu_{\eta_{\gamma}}$ & $108.886$ ($1.021$) & $<0.0001$ & $94.486$ ($1.022$) & $<0.0001$ & $53.609$ ($0.584$) & $<0.0001$ \\
$\mu_{\eta_{2}}$ & $0.703$ ($0.016$) & $<0.0001$ & $0.782$ ($0.018$) & $<0.0001$ & $0.574$ ($0.013$) & $<0.0001$ \\
$\gamma$ & $93.378$ ($0.317$) & $<0.0001$ & $99.006$ ($0.408$) & $<0.0001$ & $99.160$ ($0.252$) & $<0.0001$ \\
\hline
\hline
& \multicolumn{6}{c}{\textbf{Growth Factor (Unexplained) Variances}} \\
\hline
\textbf{Para.} & \multicolumn{2}{c}{\textbf{Reading Ability}} & \multicolumn{2}{c}{\textbf{Mathematics Ability}} \\
\hline
& \textbf{Est. (SE)} & \textbf{P value} & \textbf{Est. (SE)} & \textbf{P value} \\
\hline
$\psi_{11}$ & $0.190$ ($0.021$) & $<0.0001$ & $0.048$ ($0.007$) & $<0.0001$ & $0.014$ ($0.005$) & $0.0051$ \\
$\psi_{\gamma\gamma}$ & $319.694$ ($23.975$) & $<0.0001$ & $86.583$ ($7.763$) & $<0.0001$ & $34.078$ ($3.654$) & $<0.0001$ \\
$\psi_{22}$ & $0.045$ ($0.006$) & $<0.0001$ & $0.031$ ($0.006$) & $<0.0001$ & $0.008$ ($0.006$) & $0.1824$ \\
\hline
\hline
& \multicolumn{6}{c}{\textbf{Direct Effects}} \\
\hline
\textbf{Para.} & \multicolumn{6}{c}{\textbf{Growth Factors of Reading Ability to Those of Mathematics Ability}} \\
\hline
& \textbf{Est. (SE)} & \textbf{P value} & \textbf{---} & \textbf{---} & \textbf{---} & \textbf{---} \\
\hline
$\eta^{[x]}_{1}\rightarrow{\eta^{[m]}_{1}}$ & $0.472$ ($0.043$) & $<0.0001$ & --- & --- & --- & --- \\
$\eta^{[x]}_{1}\rightarrow{\eta^{[m]}_{\gamma}}$ & $4.569$ ($1.650$) & $0.0056$ & --- & --- & --- & ---  \\
$\eta^{[x]}_{1}\rightarrow{\eta^{[m]}_{2}}$ & $-0.020$ ($0.082$) & $0.8073$ & --- & --- & --- & --- \\
$\eta^{[x]}_{\gamma}\rightarrow{\eta^{[m]}_{\gamma}}$ & $0.695$ ($0.028$) & $<0.0001$ & --- & --- & --- & --- \\
$\eta^{[x]}_{\gamma}\rightarrow{\eta^{[m]}_{2}}$ & $0.003$ ($0.003$) & $0.3173$ & --- & --- & --- & --- \\
$\eta^{[x]}_{2}\rightarrow{\eta^{[m]}_{2}}$ & $0.775$ ($0.213$) & $0.0003$ & --- & --- & --- & --- \\
\hline
& \multicolumn{6}{c}{\textbf{Growth Factors of Reading/Mathematics Ability to Those of Science Ability}} \\
\hline
\textbf{Para.} & \multicolumn{2}{c}{\textbf{Reading Ability}} & \multicolumn{2}{c}{\textbf{Mathematics Ability}} & & \\
\hline
& \textbf{Est. (SE)} & \textbf{P value} & \textbf{Est. (SE)} & \textbf{P value} & \textbf{---} & \textbf{---} \\
\hline
$\eta^{[u]}_{1}\rightarrow{\eta^{[y]}_{1}}$ & $0.210$ ($0.052$) & $0.0001$ & $0.274$ ($0.075$) & $0.0003$ & --- & --- \\
$\eta^{[u]}_{1}\rightarrow{\eta^{[y]}_{\gamma}}$ & $9.455$ ($2.013$) & $<0.0001$ & $-8.667$ ($3.662$) & $0.0179$ & --- & --- \\
$\eta^{[u]}_{1}\rightarrow{\eta^{[y]}_{2}}$ & $-0.314$ ($0.107$) & $0.0033$ & $-0.049$ ($0.134$) & $0.7146$ & --- & --- \\
$\eta^{[u]}_{\gamma}\rightarrow{\eta^{[y]}_{\gamma}}$ & $0.030$ ($0.052$) & $0.5640$ & $0.434$ ($0.065$) & $<0.0001$ & --- & --- \\
$\eta^{[u]}_{\gamma}\rightarrow{\eta^{[y]}_{2}}$ & $0.010$ ($0.004$) & $0.0124$ & $0.004$ ($0.003$) & $0.1824$ & --- & --- \\
$\eta^{[u]}_{2}\rightarrow{\eta^{[y]}_{2}}$ & $0.813$ ($0.250$) & $0.0011$ & $0.454$ ($0.126$) & $0.0003$ & --- & --- \\
\hline
\hline
& \multicolumn{6}{c}{\textbf{Indirect Effects}} \\
\hline
& \textbf{---} & \textbf{---} & \textbf{---} & \textbf{---} & \textbf{Est. (SE)} & \textbf{P value} \\
\hline
$\eta^{[x]}_{1}\rightarrow{\eta^{[m]}_{1}}\rightarrow{\eta^{[y]}_{1}}$ & --- & --- & --- & --- & $0.129$ ($0.036$) & $0.0003$ \\
$\eta^{[x]}_{1}\rightarrow{\eta^{[m]}_{1}}\rightarrow{\eta^{[y]}_{\gamma}}$ & --- & --- & --- & --- & $-4.092$ ($1.822$) & $0.0247$ \\
$\eta^{[x]}_{1}\rightarrow{\eta^{[m]}_{1}}\rightarrow{\eta^{[y]}_{2}}$ & --- & --- & --- & --- & $-0.023$ ($0.063$) & $0.7151$ \\
$\eta^{[x]}_{1}\rightarrow{\eta^{[m]}_{\gamma}}\rightarrow{\eta^{[y]}_{\gamma}}$ & --- & --- & --- & --- & $1.985$ ($0.767$) & $0.0097$ \\
$\eta^{[x]}_{1}\rightarrow{\eta^{[m]}_{\gamma}}\rightarrow{\eta^{[y]}_{2}}$ & --- & --- & --- & --- & $0.016$ ($0.014$) & $0.2531$ \\
$\eta^{[x]}_{1}\rightarrow{\eta^{[m]}_{2}}\rightarrow{\eta^{[y]}_{2}}$ & --- & --- & --- & --- & $-0.009$ ($0.036$) & $0.8026$ \\
$\eta^{[x]}_{\gamma}\rightarrow{\eta^{[m]}_{\gamma}}\rightarrow{\eta^{[y]}_{\gamma}}$ & --- & --- & --- & --- & $0.302$ ($0.047$) & $<0.0001$ \\
$\eta^{[x]}_{\gamma}\rightarrow{\eta^{[m]}_{\gamma}}\rightarrow{\eta^{[y]}_{2}}$ & --- & --- & --- & --- & $0.002$ ($0.002$) & $0.3173$ \\
$\eta^{[x]}_{\gamma}\rightarrow{\eta^{[m]}_{2}}\rightarrow{\eta^{[y]}_{2}}$ & --- & --- & --- & --- & $0.002$ ($0.001$) & $0.0455$ \\
$\eta^{[x]}_{2}\rightarrow{\eta^{[m]}_{2}}\rightarrow{\eta^{[y]}_{2}}$ & --- & --- & --- & --- & $0.351$ ($0.089$) & $0.0001$ \\
\hline
\hline
& \multicolumn{6}{c}{\textbf{Total Effects}} \\
\hline
& \textbf{---} & \textbf{---} & \textbf{---} & \textbf{---} & \textbf{Est. (SE)} & \textbf{P value} \\
\hline
$\eta^{[x]}_{1}\rightarrow{\eta^{[y]}_{1}}$ & --- & --- & --- & --- & $0.339$ ($0.036$) & $<0.0001$ \\
$\eta^{[x]}_{1}\rightarrow{\eta^{[y]}_{\gamma}}$ & --- & --- & --- & --- & $7.347$ ($1.325$) & $<0.0001$ \\
$\eta^{[x]}_{1}\rightarrow{\eta^{[y]}_{2}}$ & --- & --- & --- & --- & $-0.330$ ($0.099$) & $0.0009$ \\
$\eta^{[x]}_{\gamma}\rightarrow{\eta^{[y]}_{\gamma}}$ & --- & --- & --- & --- & $0.332$ ($0.023$) & $<0.0001$ \\
$\eta^{[x]}_{\gamma}\rightarrow{\eta^{[y]}_{2}}$ & --- & --- & --- & --- & $0.015$ ($0.003$) & $<0.0001$ \\
$\eta^{[x]}_{2}\rightarrow{\eta^{[y]}_{2}}$ & --- & --- & --- & --- & $1.164$ ($0.243$) & $<0.0001$ \\
\hline
\hline
\end{tabular}
\label{tbl:Est2}
\end{threeparttable}
\end{table}

\bibliographystyle{apalike}
\bibliography{Extension4}

\begin{thebibliography}{}

\bibitem[Akbaşlı et~al., 2016]{Akba2016Reading}
Akbaşlı, S., Şahin, M., and Yaykiran, Z. (2016).
\newblock The effect of reading comprehension on the performance in science and
  mathematics.
\newblock {\em Journal of Education and Practice}, 7:108--121.

\bibitem[Baron and Kenny, 1986]{Baron1986indirect}
Baron, R.~M. and Kenny, D.~A. (1986).
\newblock The moderator-mediator variable distinction in social psychological
  research: conceptual, strategic, and statistical considerations.
\newblock {\em Journal of personality and social psychology},
  52(6):1173–1182.

\bibitem[Blozis, 2004]{Blozis2004MGM}
Blozis, S.~A. (2004).
\newblock Structured latent curve models for the study of change in
  multivariate repeated measures.
\newblock {\em Psychological Methods}, 9(3):334--353.

\bibitem[Blozis and Cho, 2008]{Blozis2008coding}
Blozis, S.~A. and Cho, Y. (2008).
\newblock Coding and centering of time in latent curve models in the presence
  of interindividual time heterogeneity.
\newblock {\em Structural Equation Modeling: A Multidisciplinary Journal},
  15(3):413--433.

\bibitem[Blozis et~al., 2008]{Blozis2008MGM}
Blozis, S.~A., Harring, J.~R., and Mels, G. (2008).
\newblock Using lisrel to fit nonlinear latent curve models.
\newblock {\em Structural Equation Modeling: A Multidisciplinary Journal},
  15(2):346--369.

\bibitem[Boker et~al., 2020]{User2020OpenMx}
Boker, S.~M., Neale, M.~C., Maes, H.~H., Wilde, M.~J., Spiegel, M., Brick,
  T.~R., Estabrook, R., Bates, T.~C., Mehta, P., {von Oertzen}, T., Gore,
  R.~J., Hunter, M.~D., Hackett, D.~C., Karch, J., Brandmaier, A.~M., Pritikin,
  J.~N., Zahery, M., and Kirkpatrick, R.~M. (2020).
\newblock {\em OpenMx 2.17.2 User Guide}.

\bibitem[Bollen and Curran, 2005]{Bollen2005LCM}
Bollen, K.~A. and Curran, P.~J. (2005).
\newblock {\em Latent Curve Models: A Structural Equation Perspective}.
\newblock John Wiley \& Sons, Inc.

\bibitem[Cheong et~al., 2003]{Cheong2003mediate}
Cheong, J., Mackinnon, D.~P., and Khoo, S.~T. (2003).
\newblock Investigation of mediational processes using parallel process latent
  growth curve modeling.
\newblock {\em Structural equation modeling : a multidisciplinary journal},
  10(2):238--262.

\bibitem[Cheung and Lau, 2008]{Cheung2008Mediate}
Cheung, G.~W. and Lau, R.~S. (2008).
\newblock Testing mediation and suppression effects of latent variables:
  Bootstrapping with structural equation models.
\newblock {\em Organizational Research Methods}, 11(2):296–325.

\bibitem[Cheung, 2009]{Cheung2009indirect}
Cheung, M. W.~L. (2009).
\newblock Comparison of methods for constructing confidence intervals of
  standardized indirect effects.
\newblock {\em Behavior Research Methods}, 41:425–438.

\bibitem[Cohen, 1988]{Cohen1988R}
Cohen, J. (1988).
\newblock {\em Statistical Power Analysis for the Behavioral Sciences (2nd
  Ed.)}.
\newblock Lawrence Erlbaum Associates.

\bibitem[Cole and Maxwell, 2003]{Cole2003mediate}
Cole, D.~A. and Maxwell, S.~E. (2003).
\newblock Testing mediational models with longitudinal data: questions and tips
  in the use of structural equation modeling.
\newblock {\em Journal of abnormal psychology}, 112(4):558–577.

\bibitem[Coulombe et~al., 2015]{Coulombe2015ignoring}
Coulombe, P., Selig, J.~P., and Delaney, H.~D. (2015).
\newblock Ignoring individual differences in times of assessment in growth
  curve modeling.
\newblock {\em International Journal of Behavioral Development}, 40(1):76--86.

\bibitem[Cudeck and du~Toit, 2003]{Cudeck2003knot_F}
Cudeck, R. and du~Toit, S. H.~C. (2003).
\newblock Nonlinear multilevel models for repeated measures data.
\newblock In Reise, S.~P. and Duan, N., editors, {\em Multilevel Modeling :
  Methodological Advances, Issues, and Applications.}, Multivariate
  Applications Book Series, chapter~2, pages 1--24. Psychology Press.

\bibitem[Dominicus et~al., 2008]{Dominicus2008knot_B}
Dominicus, A., Ripatti, S., Pedersen, N.~L., and Palmgren, J. (2008).
\newblock A random change point model for assessing variability in repeated
  measures of cognitive function.
\newblock {\em Statistics in Medicine}, 27(27):5786--5798.

\bibitem[Dumenci et~al., 2019]{Dumenci2019knee}
Dumenci, L., Perera, R.~A., Keefe, F.~J., Ang, D.~C., J., S., Jensen, M.~P.,
  and Riddle, D.~L. (2019).
\newblock Model-based pain and function outcome trajectory types for patients
  undergoing knee arthroplasty: a secondary analysis from a randomized clinical
  trial.
\newblock {\em Osteoarthritis and cartilage}, 27(6):878--884.

\bibitem[Flora, 2008]{Flora2008knot}
Flora, D.~B. (2008).
\newblock Specifying piecewise latent trajectory models for longitudinal data.
\newblock {\em Structural Equation Modeling: A Multidisciplinary Journal},
  15(3):513--533.

\bibitem[Gollob and Reichardt, 1987]{Gollob1987mediate}
Gollob, H. and Reichardt, C. (1987).
\newblock Taking account of time lags in causal models.
\newblock {\em Child Development}, 58(1):80--92.

\bibitem[Grimm et~al., 2016]{Grimm2016growth}
Grimm, K.~J., Ram, N., and Estabrook, R. (2016).
\newblock {\em Growth Modeling: Structural Equation and Multilevel Modeling
  Approaches}.
\newblock Guilford Press.

\bibitem[Harring et~al., 2006]{Harring2006nonlinear}
Harring, J.~R., Cudeck, R., and du~Toit, S. H.~C. (2006).
\newblock Fitting partially nonlinear random coefficient models as sems.
\newblock {\em Multivariate Behavioral Research}, 41(4):579--596.

\bibitem[Hayes, 2009]{Hayes2009Mediate}
Hayes, A.~F. (2009).
\newblock Beyond baron and kenny: Statistical mediation analysis in the new
  millennium.
\newblock {\em Communication Monographs}, 76(4):408–420.

\bibitem[Hoerl and Kennard, 1970a]{Hoerl1970Ridge2}
Hoerl, A.~E. and Kennard, R.~W. (1970a).
\newblock Ridge regression: Applications to nonorthogonal problems.
\newblock {\em Technometrics}, 12(1):69--82.

\bibitem[Hoerl and Kennard, 1970b]{Hoerl1970Ridge1}
Hoerl, A.~E. and Kennard, R.~W. (1970b).
\newblock Ridge regression: Biased estimation for nonorthogonal problems.
\newblock {\em Technometrics}, 12(1):55--67.

\bibitem[Hunter, 2018]{Hunter2018OpenMx}
Hunter, M.~D. (2018).
\newblock State space modeling in an open source, modular, structural equation
  modeling environment.
\newblock {\em Structural Equation Modeling: A Multidisciplinary Journal},
  25(2):307--324.

\bibitem[Kohli, 2011]{Kohli2011PLGC}
Kohli, N. (2011).
\newblock {\em Estimating unknown knots in piecewise linear-linear latent
  growth mixture models}.
\newblock PhD thesis, University of Maryland.

\bibitem[Kohli and Harring, 2013]{Kohli2013PLGC2}
Kohli, N. and Harring, J.~R. (2013).
\newblock Modeling growth in latent variables using a piecewise function.
\newblock {\em Multivariate Behavioral Research}, 48(3):370--397.

\bibitem[Kohli et~al., 2013]{Kohli2013PLGC1}
Kohli, N., Harring, J.~R., and Hancock, G.~R. (2013).
\newblock Piecewise linear-linear latent growth mixture models with unknown
  knots.
\newblock {\em Educational and Psychological Measurement}, 73(6):935--955.

\bibitem[Kohli et~al., 2015]{Kohli2015PLGC1}
Kohli, N., Hughes, J., Wang, C., Zopluoglu, C., and Davison, M.~L. (2015).
\newblock Fitting a linear-linear piecewise growth mixture model with unknown
  knots: A comparison of two common approaches to inference.
\newblock {\em Psychological Methods}, 20(2):259--275.

\bibitem[Kwok et~al., 2010]{Kwok2010simu}
Kwok, O., Luo, W., and West, S.~G. (2010).
\newblock Using modification indexes to detect turning points in longitudinal
  data: A monte carlo study.
\newblock {\em Structural Equation Modeling: A Multidisciplinary Journal},
  17(2):216--240.

\bibitem[L{\^e} et~al., 2011]{Le2011ECLS}
L{\^e}, T., Norman, G., Tourangeau, K., Brick, J.~M., and Mulligan, G. (2011).
\newblock Early childhood longitudinal study: Kindergarten class of 2010-2011
  – sample design issues.
\newblock In {\em JSM Proceedings 2011}, pages 1629--1639. Alexandria, VA:
  American Statistical Association.

\bibitem[Liu, 2019]{Liu2019knot}
Liu, J. (2019).
\newblock {\em Estimating Knots in Bilinear Spline Growth Models with
  Time-invariant Covariates in the Framework of Individual Measurement
  Occasions}.
\newblock PhD thesis, Virginia Commonwealth University.

\bibitem[Liu and Perera, 2021]{Liu2021PBLSGM}
Liu, J. and Perera, R.~A. (2021).
\newblock Estimating knots and their association in parallel bilinear spline
  growth curve models in the framework of individual measurement occasions.
\newblock {\em Psychological Methods (Advance online publication)}.

\bibitem[Liu et~al., 2019a]{Liu2019BLSGM}
Liu, J., Perera, R.~A., Kang, L., Kirkpatrick, R.~M., and Sabo, R.~T. (2019a).
\newblock Obtaining interpretable parameters from reparameterized longitudinal
  models: transformation matrices between growth factors in two
  parameter-spaces.

\bibitem[Liu et~al., 2019b]{Liu2019BLSGMM}
Liu, J., Perera, R.~A., Kang, L., Sabo, R.~T., and Kirkpatrick, R.~M. (2019b).
\newblock Hybridizing two-step growth mixture model and exploratory factor
  analysis to examine heterogeneity in nonlinear trajectories.

\bibitem[Lock et~al., 2018]{Lock2018knot_B}
Lock, E.~F., Kohli, N., and Bose, M. (2018).
\newblock Detecting multiple random changepoints in bayesian piecewise growth
  mixture models.
\newblock {\em Psychometrika}, 83(3):733--750.

\bibitem[MacKinnon, 2008]{MacKinnon2008mediate}
MacKinnon, D.~P. (2008).
\newblock {\em Introduction to statistical mediation analysis.}
\newblock Taylor \& Francis Group/Lawrence Erlbaum Associates., New York,
  fourth edition.

\bibitem[MacKinnon et~al., 2000]{MacKinnon2000Mediator}
MacKinnon, D.~P., Krull, J.~L., and Lockwood, C.~M. (2000).
\newblock Equivalence of the mediation, confounding and suppression effect.
\newblock {\em Prevention Science}, 1:173–181.

\bibitem[MacKinnon et~al., 2002]{MacKinnon2002Mediator}
MacKinnon, D.~P., Lockwood, C.~M., Hoffman, J.~M., West, S.~G., and Sheets, V.
  (2002).
\newblock A comparison of methods to test mediation and other intervening
  variable effects.
\newblock {\em Psychological Methods}, 7(1):83–104.

\bibitem[Maxwell and Cole, 2007]{Maxwell2007mediate}
Maxwell, S.~E. and Cole, D.~A. (2007).
\newblock Bias in cross-sectional analyses of longitudinal mediation.
\newblock {\em Psychological methods}, 12(1):23–44.

\bibitem[McArdle and Nesselroade, 1994]{McArdle1994LCSM}
McArdle, J.~J. and Nesselroade, J.~R. (1994).
\newblock Using multivariate data to structure developmental change.
\newblock In Cohen, S.~H. and Reese, H.~W., editors, {\em Life-span
  developmental psychology: Methodological contributions}, page 223–267.
  Lawrence Erlbaum Associates, Inc.

\bibitem[McArdle and Wang, 2008]{McArdle2008knot_B}
McArdle, J.~J. and Wang, L. (2008).
\newblock Modeling age-based turning points in longitudinal life-span growth
  curves of cognition.
\newblock In Cohen, P., editor, {\em Applied data analytic techniques for
  turning points research}, Multivariate applications series., chapter~2, pages
  1--24. Routledge Taylor \& Francis Group.

\bibitem[Mehta and Neale, 2005]{Mehta2005people}
Mehta, P.~D. and Neale, M.~C. (2005).
\newblock People are variables too: Multilevel structural equations modeling.
\newblock {\em Psychological Methods}, 10(3):259--284.

\bibitem[Mehta and West, 2000]{Mehta2000people}
Mehta, P.~D. and West, S.~G. (2000).
\newblock Putting the individual back into individual growth curves.
\newblock {\em Psychological Methods}, 5(1):23--43.

\bibitem[Morris et~al., 2019]{Morris2019simulation}
Morris, T.~P., White, I.~R., and Crowther, M.~J. (2019).
\newblock Using simulation studies to evaluate statistical methods.
\newblock {\em Statistics in Medicine}, 38(11):2074--2102.

\bibitem[Muniz~Terrera et~al., 2011]{Muniz2011knot_B}
Muniz~Terrera, G., van~den Hout, A., and Matthews, F.~E. (2011).
\newblock Random change point models: investigating cognitive decline in the
  presence of missing data.
\newblock {\em Journal of Applied Statistics.}, 38(4):705--716.

\bibitem[Muth\'{e}n and Muth\'{e}n, 2017]{Muthen2017Mplus}
Muth\'{e}n, L.~K. and Muth\'{e}n, B.~O. (2017).
\newblock {\em Mplus: Statistical Analysis with Latent Variables: User’s
  Guide (Version 8)}.

\bibitem[Neale et~al., 2016]{OpenMx2016package}
Neale, M.~C., Hunter, M.~D., Pritikin, J.~N., Zahery, M., Brick, T.~R.,
  Kirkpatrick, R.~M., Estabrook, R., Bates, T.~C., Maes, H.~H., and Boker,
  S.~M. (2016).
\newblock Open{M}x 2.0: {E}xtended structural equation and statistical
  modeling.
\newblock {\em Psychometrika}, 81:535--549.

\bibitem[Peralta et~al., 2020]{Peralta2020PBLSGM}
Peralta, Y., Kohli, N., Lock, E.~F., and Davison, M.~L. (2020).
\newblock Bayesian modeling of associations in bivariate piecewise linear
  mixed-effects models.
\newblock {\em Psychological Methods (Advance online publication)}.

\bibitem[Preacher and Hancock, 2015]{Preacher2015repara}
Preacher, K.~J. and Hancock, G.~R. (2015).
\newblock Meaningful aspects of change as novel random coefficients: A general
  method for reparameterizing longitudinal models.
\newblock {\em Psychological Methods}, 20(1):84--101.

\bibitem[Pritikin et~al., 2015]{Pritikin2015OpenMx}
Pritikin, J.~N., Hunter, M.~D., and Boker, S.~M. (2015).
\newblock Modular open-source software for {I}tem {F}actor {A}nalysis.
\newblock {\em Educational and Psychological Measurement}, 75(3):458--474.

\bibitem[Riddle et~al., 2015]{Riddle2015knee}
Riddle, D.~L., Perera, R.~A., Jiranek, W.~A., and Dumenci, L. (2015).
\newblock Using surgical appropriateness criteria to examine outcomes of total
  knee arthroplasty in a united states sample.
\newblock {\em Arthritis care \& research.}, 67(3):349--357.

\bibitem[Santosa and Symes, 1986]{Santosa1986LASSO}
Santosa, F. and Symes, W.~W. (1986).
\newblock Linear inversion of band-limited reflection seismograms.
\newblock {\em SIAM Journal on Scientific and Statistical Computing},
  7(4):1307--1330.

\bibitem[Selig and Preacher, 2009]{Selig2009mediate}
Selig, J.~P. and Preacher, K.~J. (2009).
\newblock Mediation models for longitudinal data in developmental research.
\newblock {\em Research in Human Development}, 6(2-3):144--164.

\bibitem[Shrout and Bolger, 2002]{Shrout2002Mediate}
Shrout, P.~E. and Bolger, N. (2002).
\newblock Mediation in experimental and nonexperimental studies: New procedures
  and recommendations.
\newblock {\em Psychological Methods}, 7(4):422–445.

\bibitem[Sobel, 1982]{Sobel1982indirect}
Sobel, M.~E. (1982).
\newblock Asymptotic confidence intervals for indirect effects in structural
  equation models.
\newblock {\em Sociological Methodology}, 13:290–312.

\bibitem[Sobel, 1986]{Sobel1986indirect}
Sobel, M.~E. (1986).
\newblock Some new results on indirect effects and their standard errors in
  covariance structural models.
\newblock {\em Sociological Methodology}, 16:159–186.

\bibitem[Soest and Hagtvet, 2011]{Soest2011mediate}
Soest, T. and Hagtvet, K.~A. (2011).
\newblock Mediation analysis in a latent growth curve modeling framework.
\newblock {\em Structural equation modeling : a multidisciplinary journal},
  18(2):289--314.

\bibitem[Sterba, 2014]{Sterba2014individually}
Sterba, S.~K. (2014).
\newblock Fitting nonlinear latent growth curve models with individually
  varying time points.
\newblock {\em Structural Equation Modeling: A Multidisciplinary Journal},
  21(4):630--647.

\bibitem[Torgesen, 2002]{Torgesen2002Reading}
Torgesen, J.~K. (2002).
\newblock The prevention of reading difficulties.
\newblock {\em Journal of School Psychology}, 40(1):7--26.

\bibitem[Tourangeau et~al., 2013]{Tourangeau2013ECLS}
Tourangeau, K., Nord, C., L{\^e}, T., Sorongon, A.~G., Hagedorn, M.~C., Daly,
  P., and Westat (2013).
\newblock {\em User’s manual for the ECLS-K:2011 Kindergarten data file and
  electronic codebook.}

\bibitem[Venables and Ripley, 2002]{Venables2002Statistics}
Venables, W.~N. and Ripley, B.~D. (2002).
\newblock {\em Modern Applied Statistics with S}.
\newblock Springer, New York, fourth edition.

\bibitem[Wang and McArdle, 2008]{Wang2008knot_B}
Wang, L. and McArdle, J.~J. (2008).
\newblock A simulation study comparison of bayesian estimation with
  conventional methods for estimating unknown change points.
\newblock {\em Structural Equation Modeling: A Multidisciplinary Journal},
  15(1):52--74.

\bibitem[Wasserstein et~al., 2019]{Wasserstein2019Pvalue}
Wasserstein, R.~L., Schirm, A.~L., and Lazar, N.~A. (2019).
\newblock Moving to a world beyond `p<0.05'.
\newblock {\em The American Statistician}, 73(sup1):1--19.

\bibitem[Zou and Hastie, 2005]{Zou2005Elastic}
Zou, H. and Hastie, T. (2005).
\newblock Regularization and variable selection via the elastic net.
\newblock {\em Journal of the Royal Statistical Society, Series B.},
  67(2):301–320.

\end{thebibliography}

\renewcommand\thefigure{\arabic{figure}}
\setcounter{figure}{0}

\begin{figure}
\centering
\includegraphics[width=0.6\textwidth]{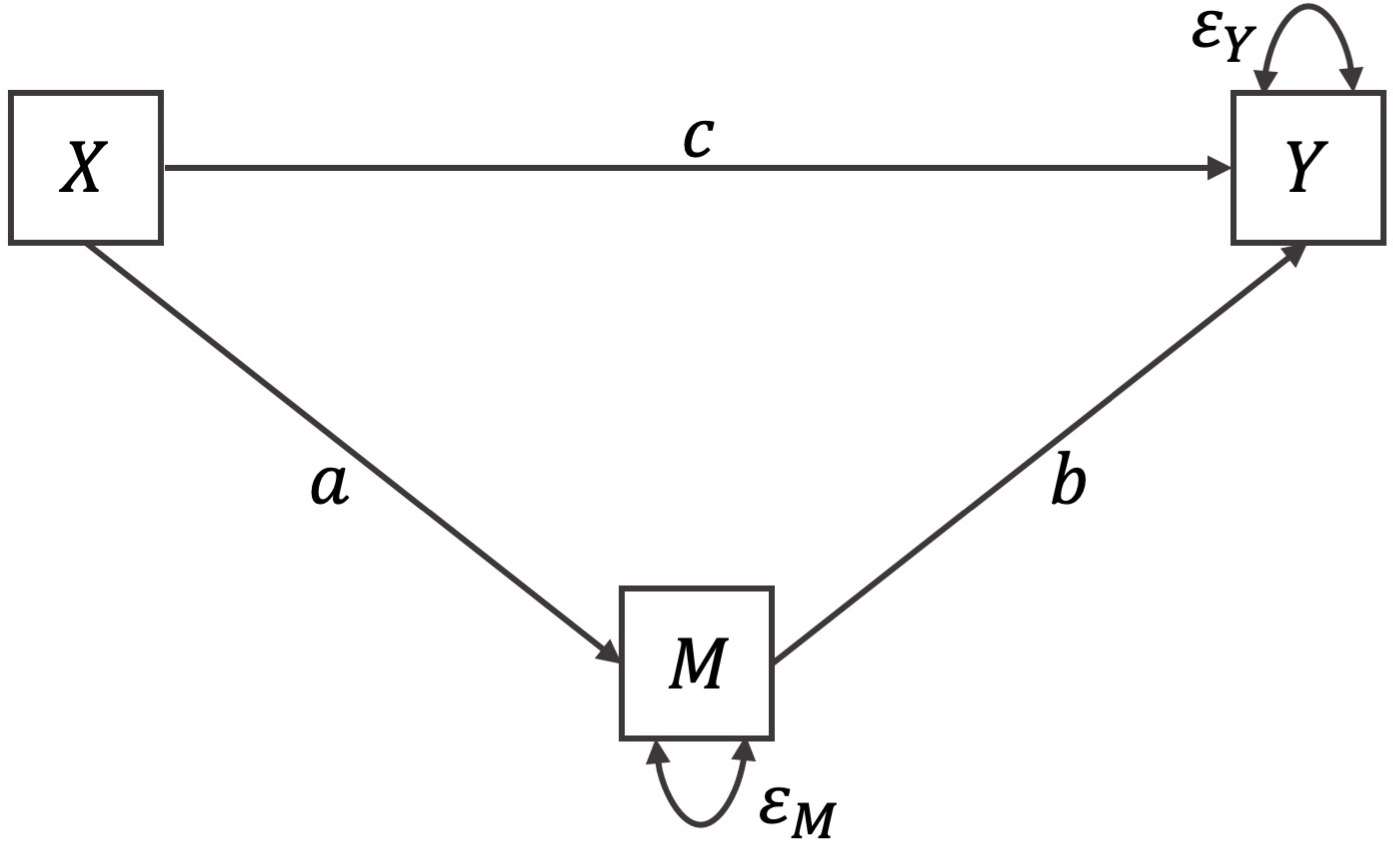}
\caption{Diagram of a Simple Mediation Model}
\label{fig:Med}
\end{figure}

\begin{figure}
\centering
\includegraphics[width=0.6\textwidth]{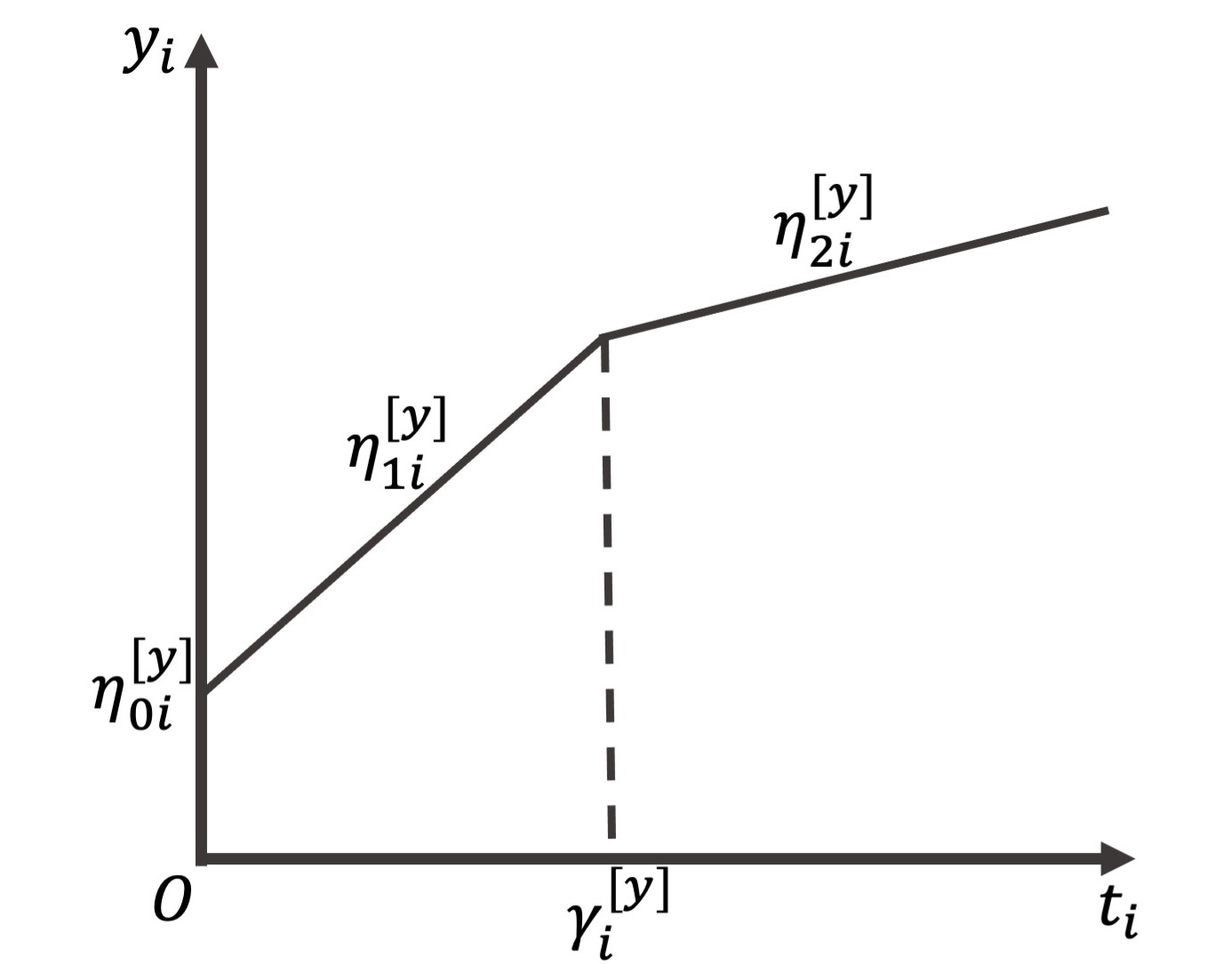}
\caption{Within-individual Change over Time with Bilinear Spline Functional Form}
\label{fig:knot}
\end{figure}

\begin{figure}
\centering
\begin{subfigure}{0.5\textwidth}
 \centering
 \includegraphics[width=\linewidth]{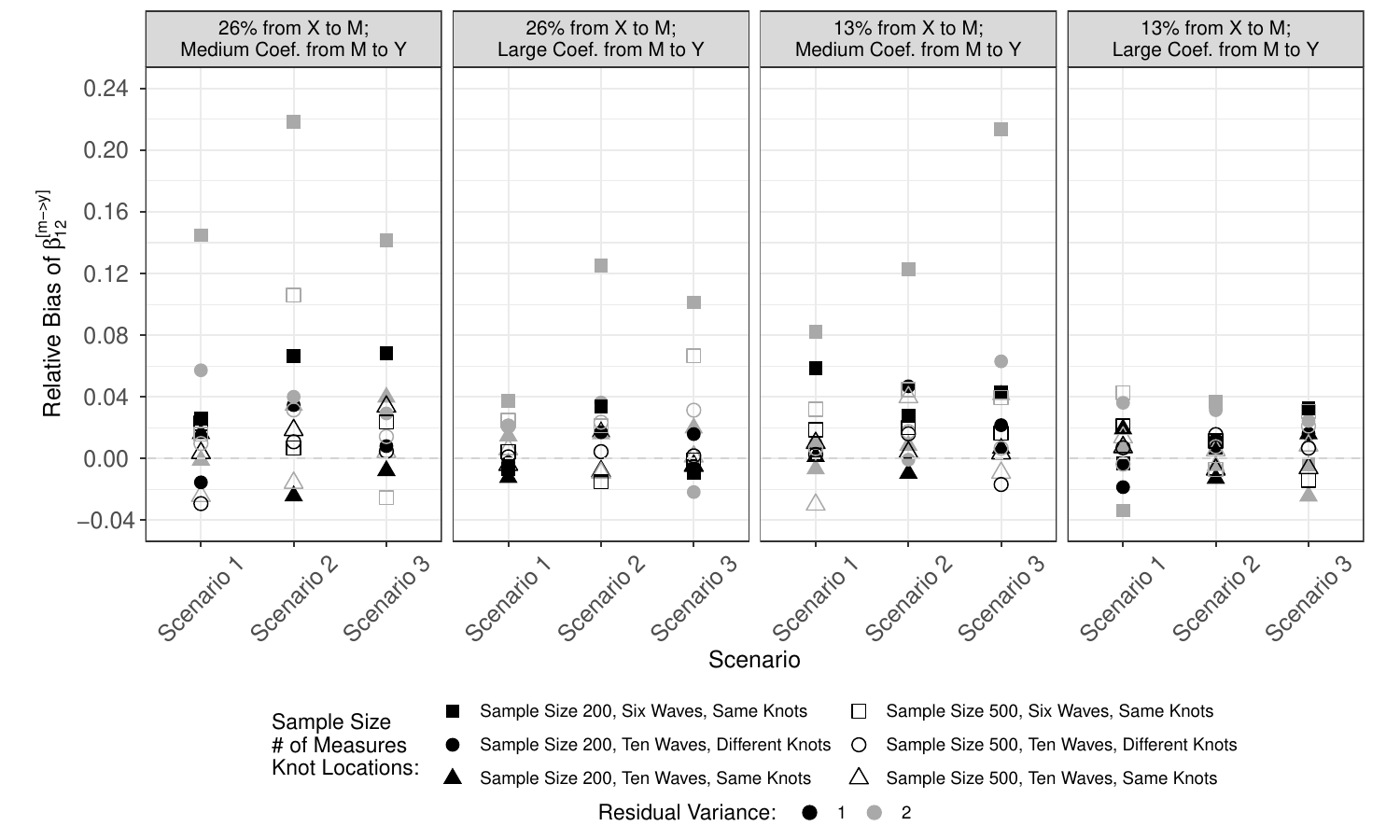}
 \caption{Relative Biases of $\beta^{[m\rightarrow{y}]}_{12}$}
 \label{fig:rBias_M1_m1y2}
\end{subfigure}%
\begin{subfigure}{0.5\textwidth}
 \centering
 \includegraphics[width=\linewidth]{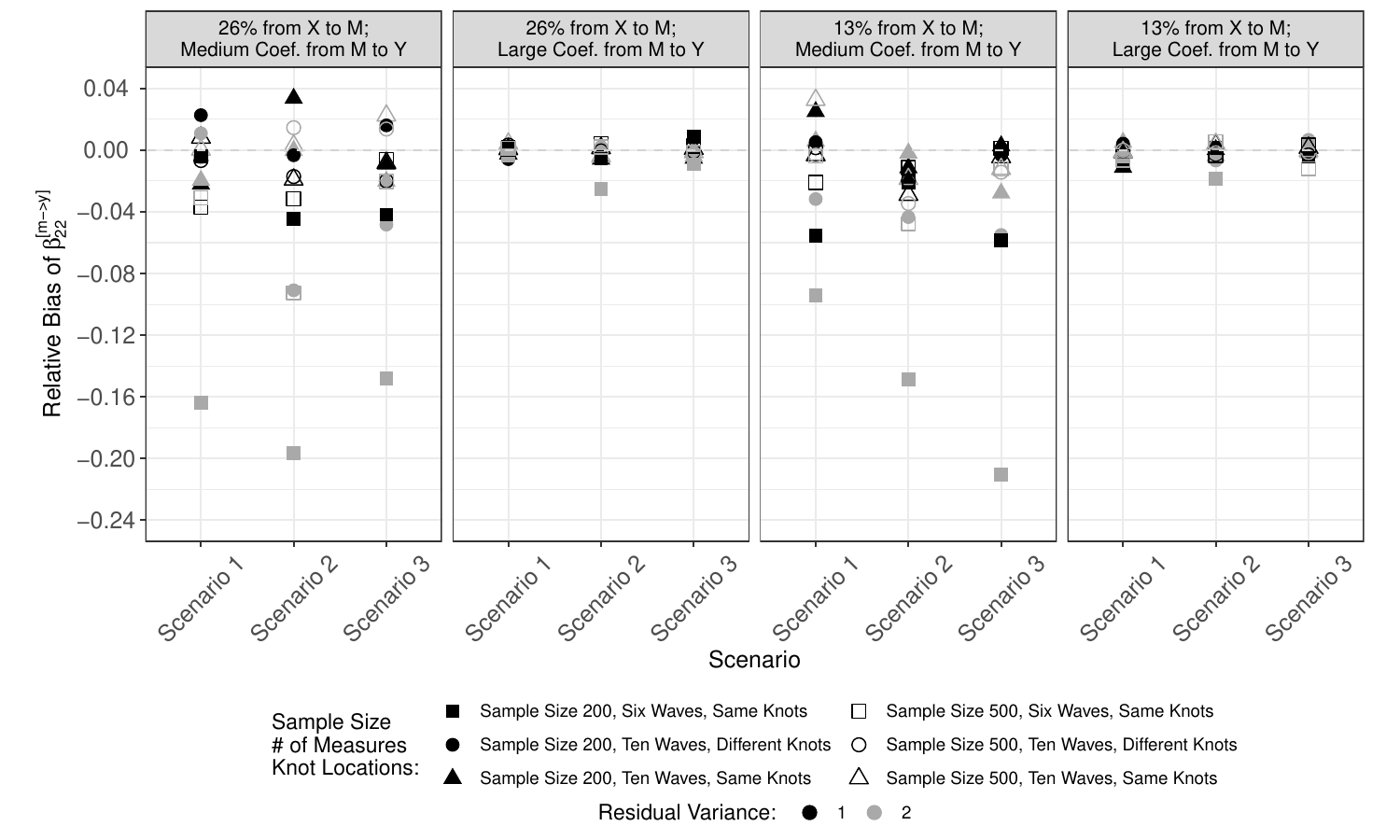}
 \caption{Relative Biases of $\beta^{[m\rightarrow{y}]}_{22}$}
 \label{fig:rBias_M1_m2y2}
\end{subfigure}
\caption{Summary of Relative Biases for Coefficients with Some Bias Greater than $10\%$ for Model 1}
\label{fig:rBias_M1_coef}
\end{figure}

\begin{figure}
\centering
\begin{subfigure}{0.5\textwidth}
 \centering
 \includegraphics[width=\linewidth]{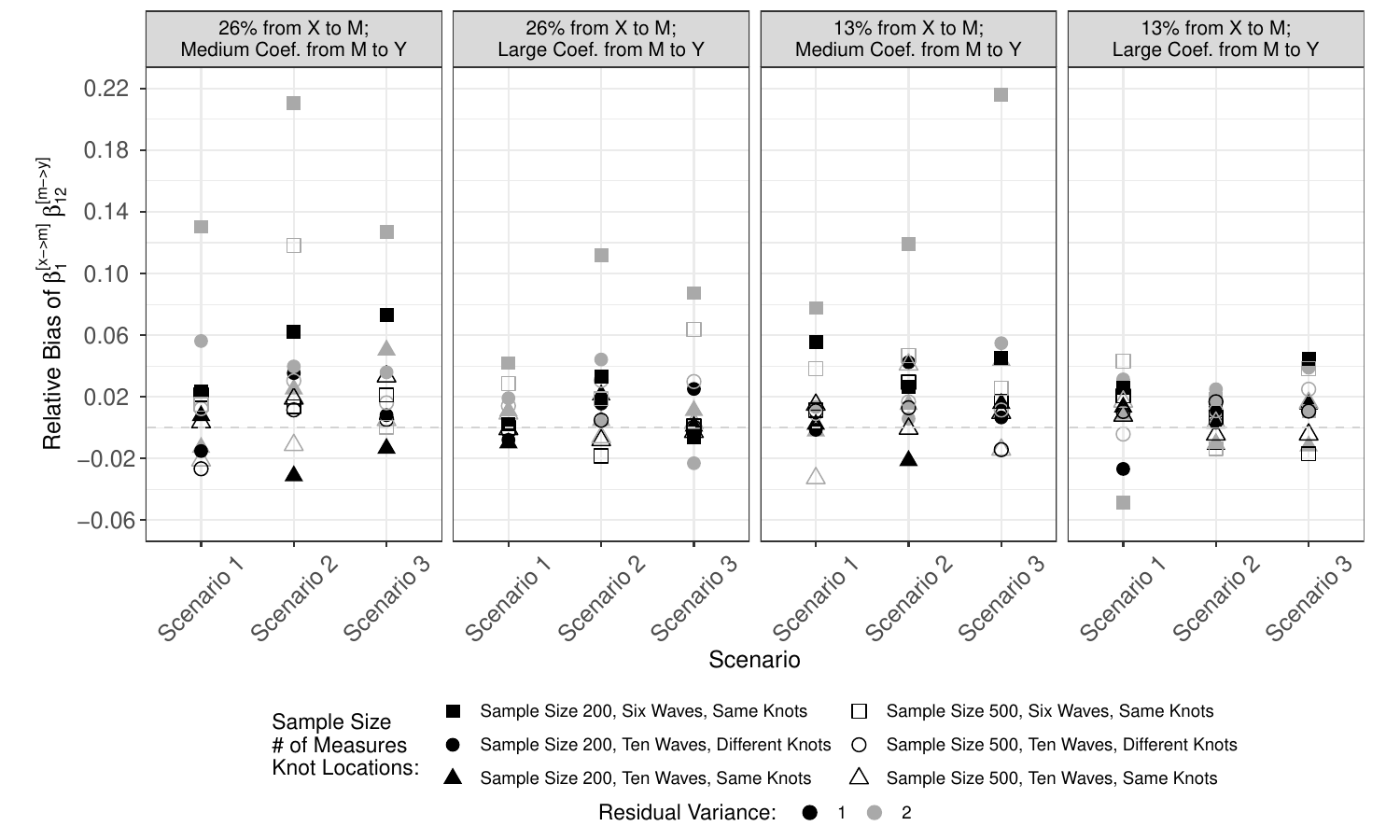}
 \caption{Relative Biases of $x\rightarrow{m_{1}\rightarrow{y_{2}}}$}
 \label{fig:rBias_M1_xm1y2}
\end{subfigure}%
\begin{subfigure}{0.5\textwidth}
 \centering
 \includegraphics[width=\linewidth]{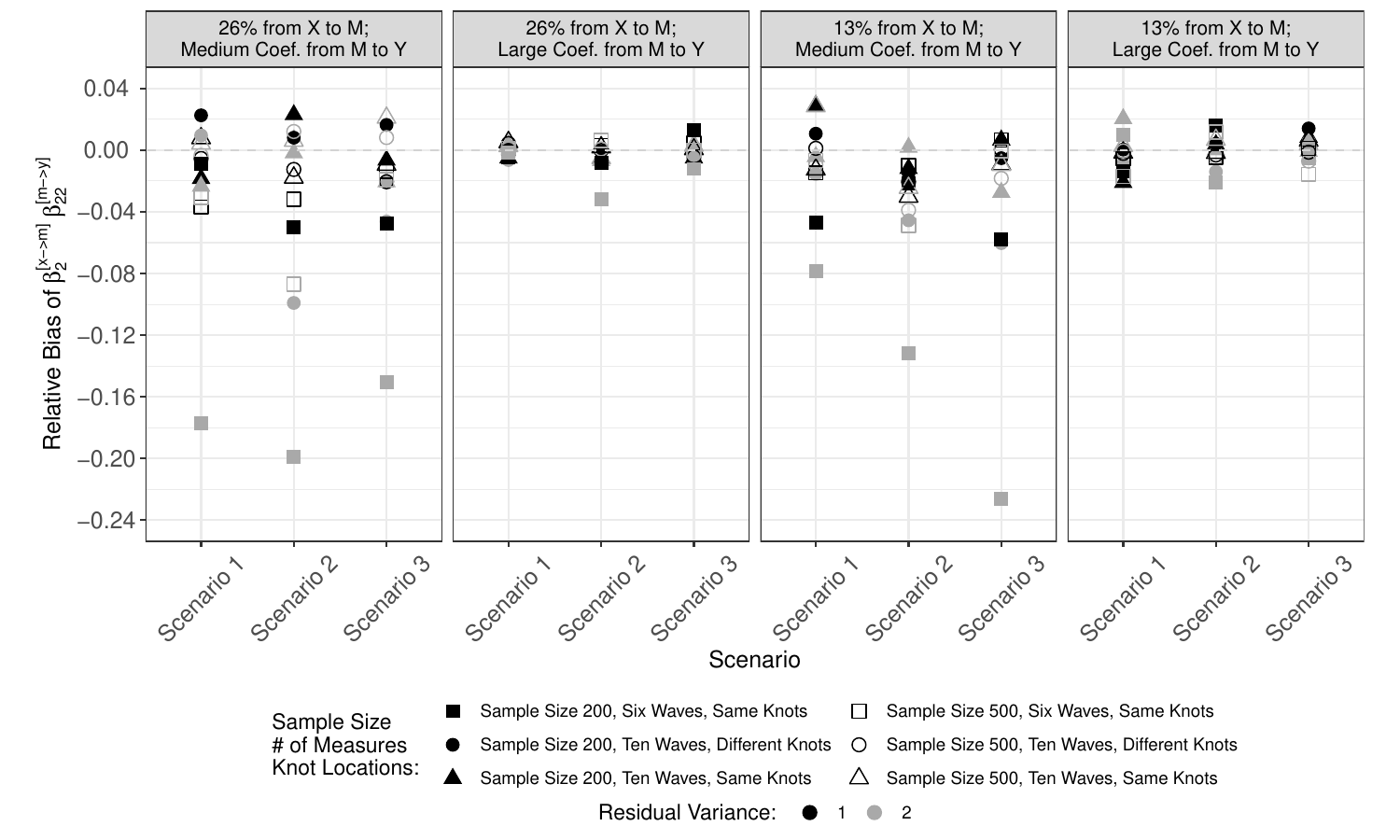}
 \caption{Relative Biases of $x\rightarrow{m_{2}\rightarrow{y_{2}}}$}
 \label{fig:rBias_M1_xm2y2}
\end{subfigure}
\caption{Summary of Relative Biases for Mediated Effects with Some Bias Greater than $10\%$ for Model 1}
\label{fig:rBias_M1_med}
\end{figure}

\begin{figure}
\centering
\begin{subfigure}{0.5\textwidth}
 \centering
 \includegraphics[width=\linewidth]{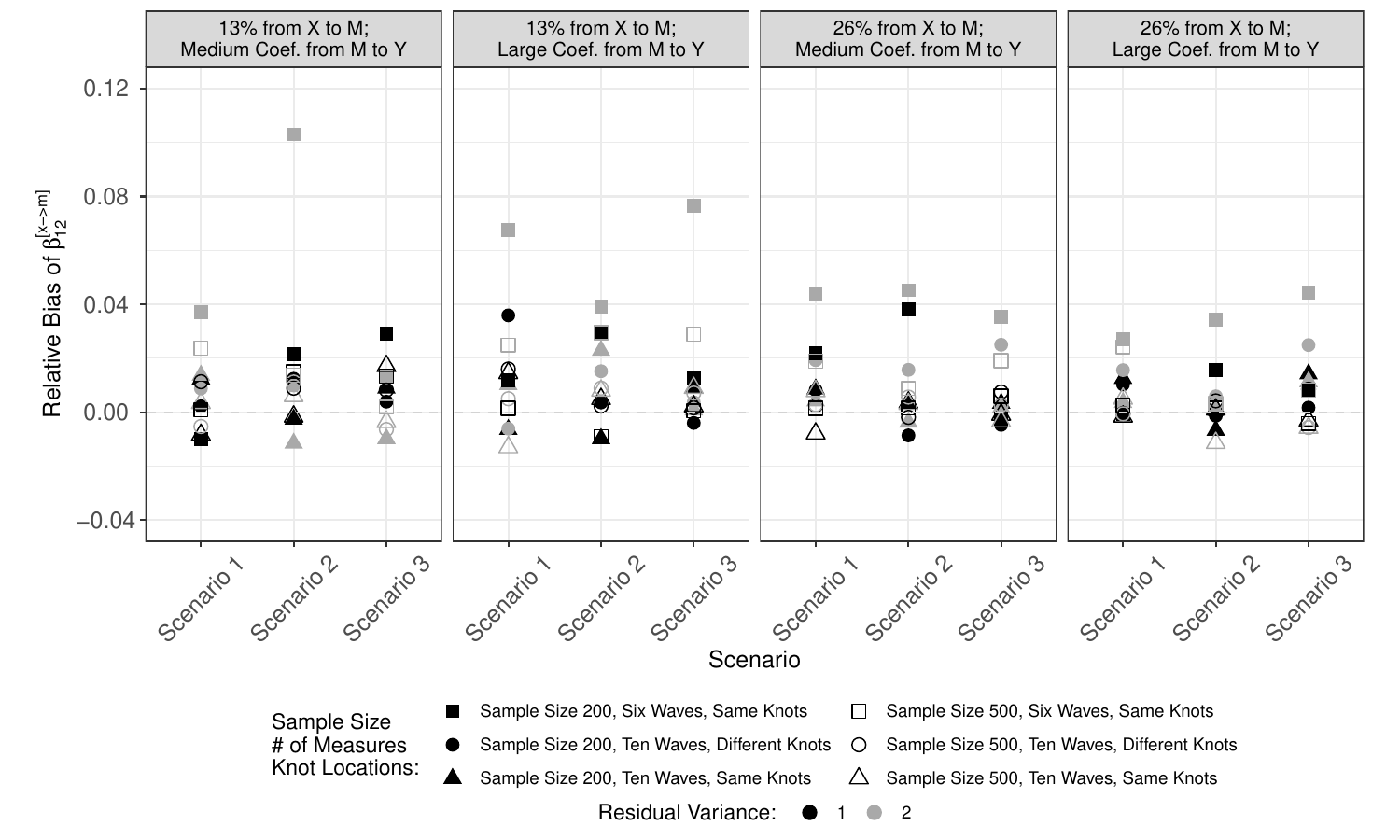}
 \caption{Relative Biases of $\beta^{[x\rightarrow{m}]}_{12}$}
 \label{fig:rBias_M2_x1m2}
\end{subfigure}%
\begin{subfigure}{0.5\textwidth}
 \centering
 \includegraphics[width=\linewidth]{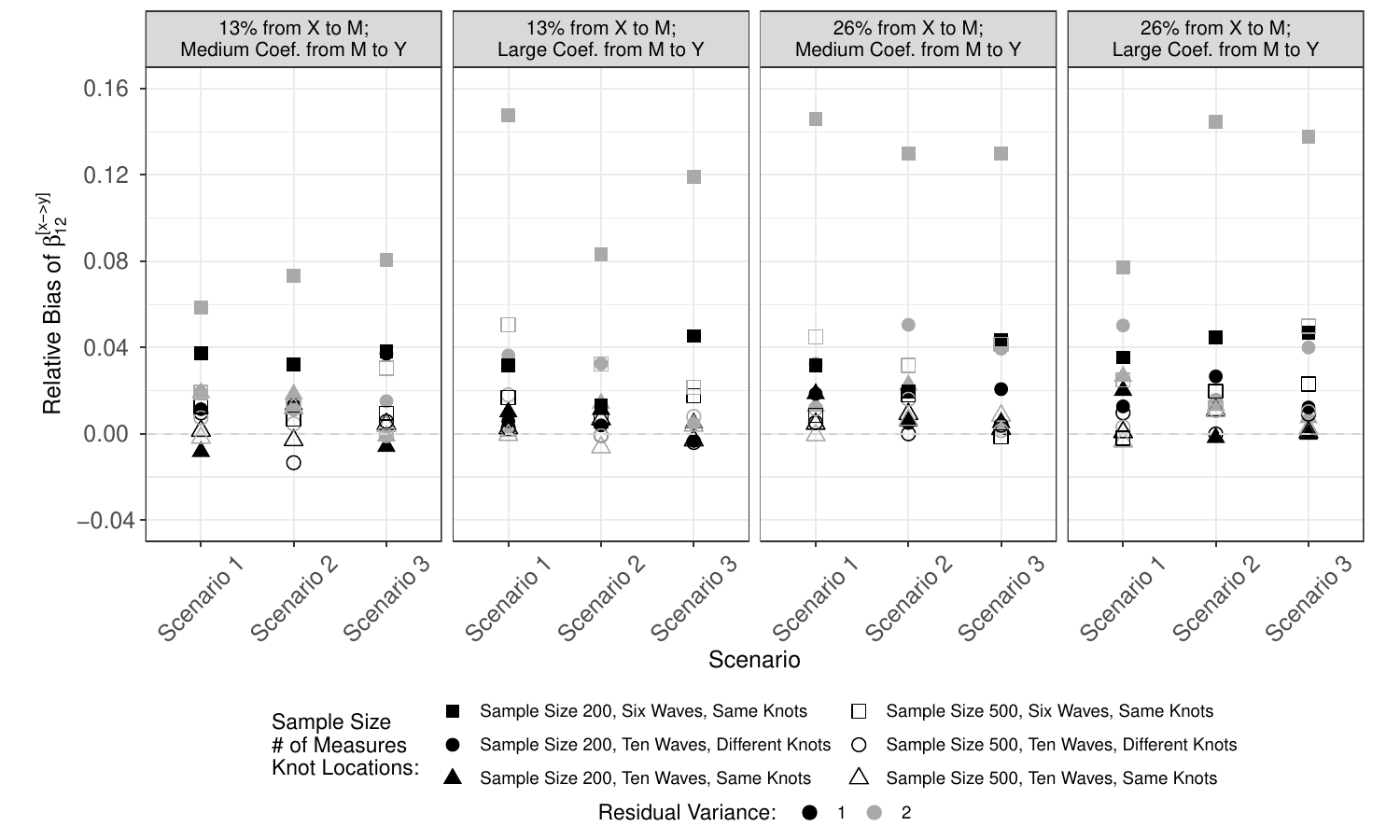}
 \caption{Relative Biases of $\beta^{[x\rightarrow{y}]}_{12}$}
 \label{fig:rBias_M2_x1y2}
\end{subfigure}
\begin{subfigure}{0.5\textwidth}
 \centering
 \includegraphics[width=\linewidth]{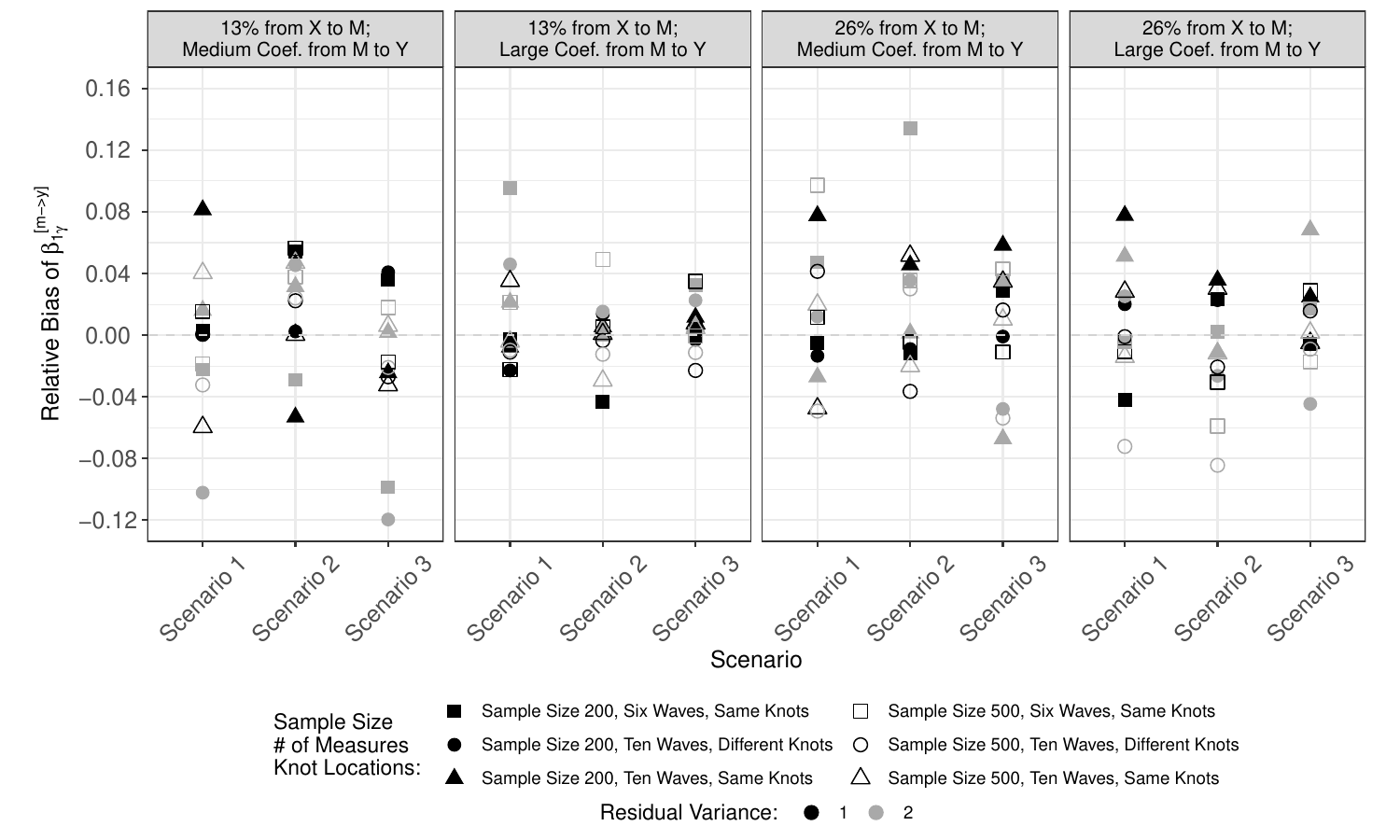}
 \caption{Relative Biases of $\beta^{[m\rightarrow{y}]}_{1\gamma}$}
 \label{fig:rBias_M2_m1yr}
\end{subfigure}%
\begin{subfigure}{0.5\textwidth}
 \centering
 \includegraphics[width=\linewidth]{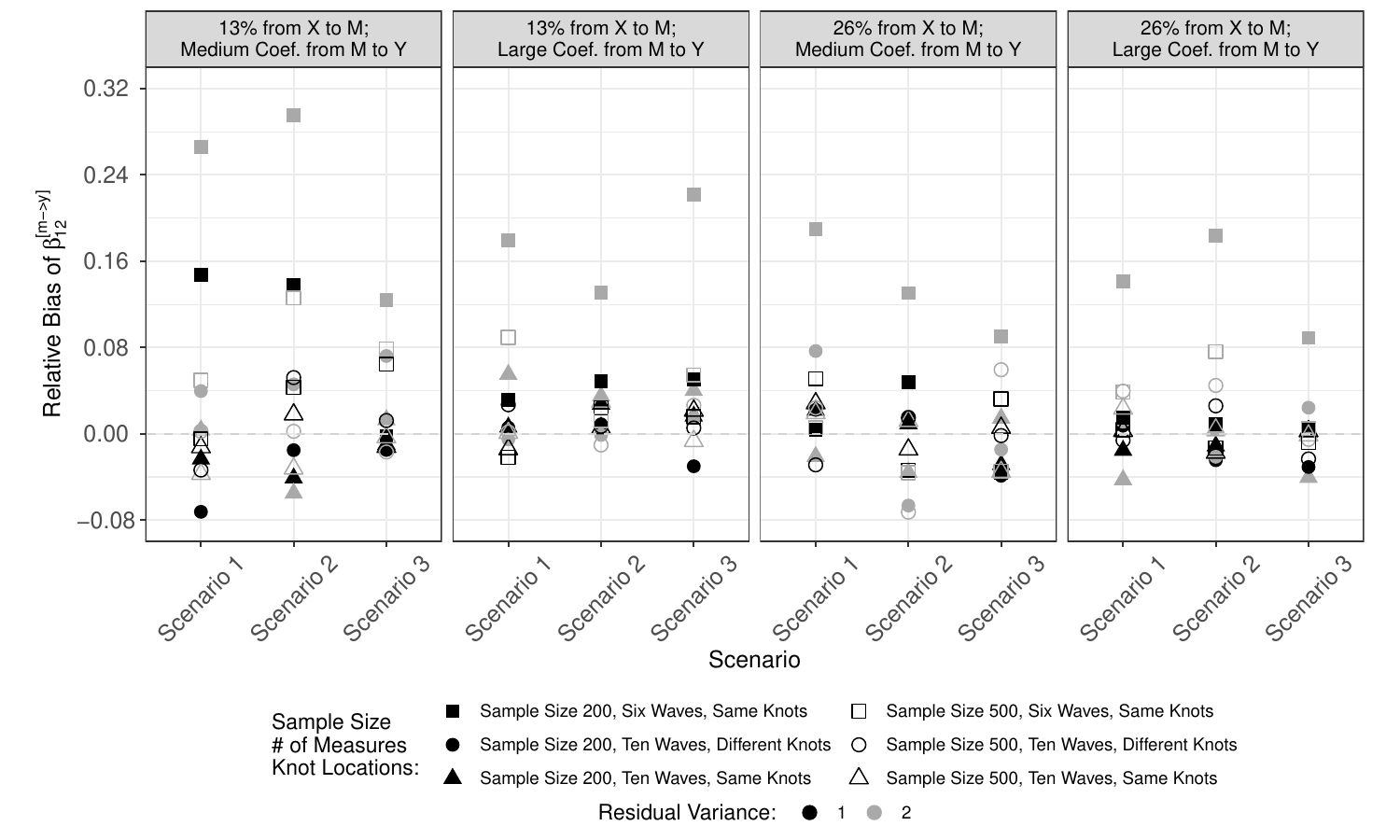}
 \caption{Relative Biases of $\beta^{[m\rightarrow{y}]}_{12}$}
 \label{fig:rBias_M2_m1y2}
\end{subfigure}
\begin{subfigure}{0.5\textwidth}
 \centering
 \includegraphics[width=\linewidth]{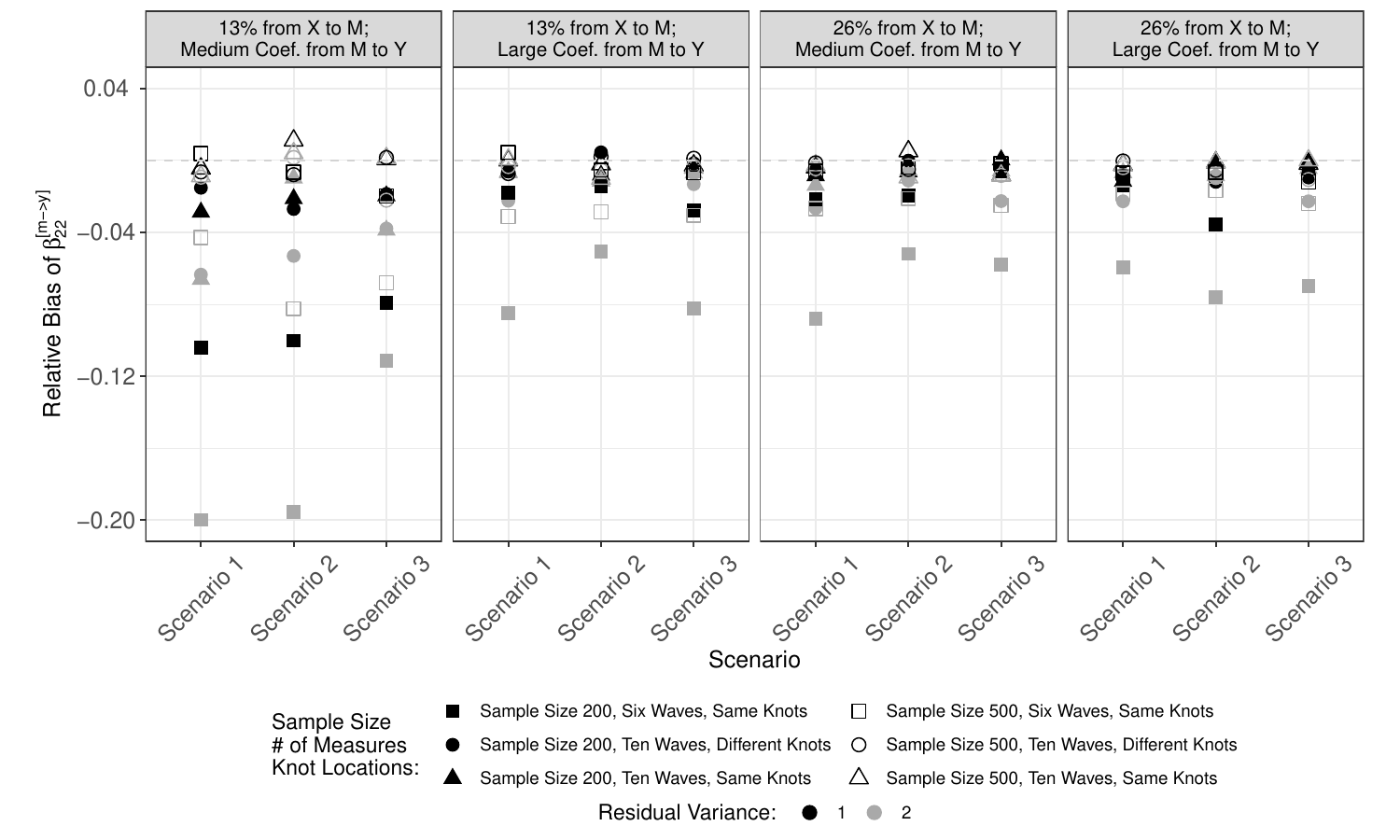}
 \caption{Relative Biases of $\beta^{[m\rightarrow{y}]}_{22}$}
 \label{fig:rBias_M2_m2y2}
\end{subfigure}%
\caption{Summary of Relative Biases for Coefficients with Some Bias Greater than $10\%$ for Model 2}
\label{fig:rBias_M2_coef}
\end{figure}

\begin{figure}
\centering
\begin{subfigure}{0.5\textwidth}
 \centering
 \includegraphics[width=\linewidth]{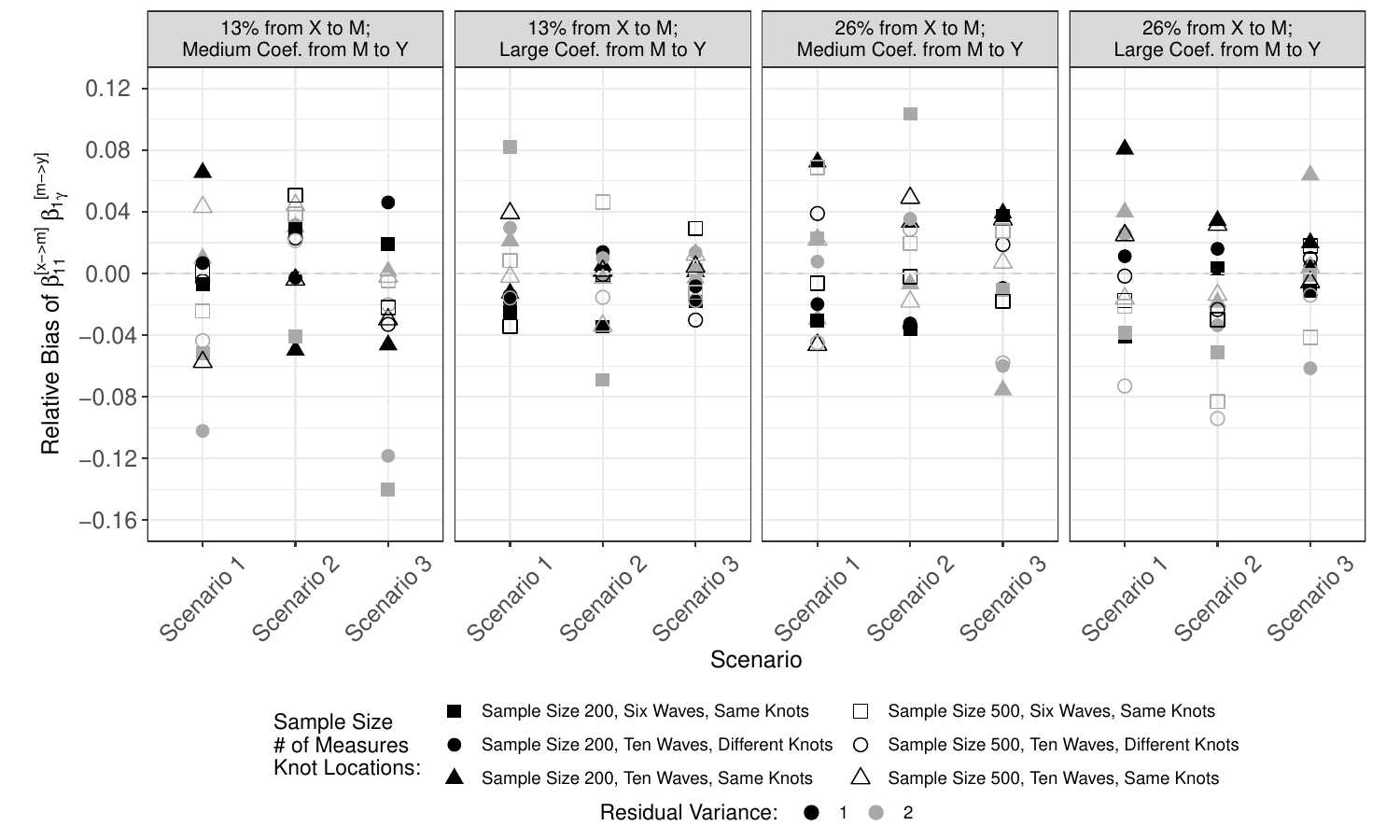}
 \caption{Relative Biases of $x_{1}\rightarrow{m_{1}\rightarrow{y_{\gamma}}}$}
 \label{fig:rBias_x1m1yr}
\end{subfigure}%
\begin{subfigure}{0.5\textwidth}
 \centering
 \includegraphics[width=\linewidth]{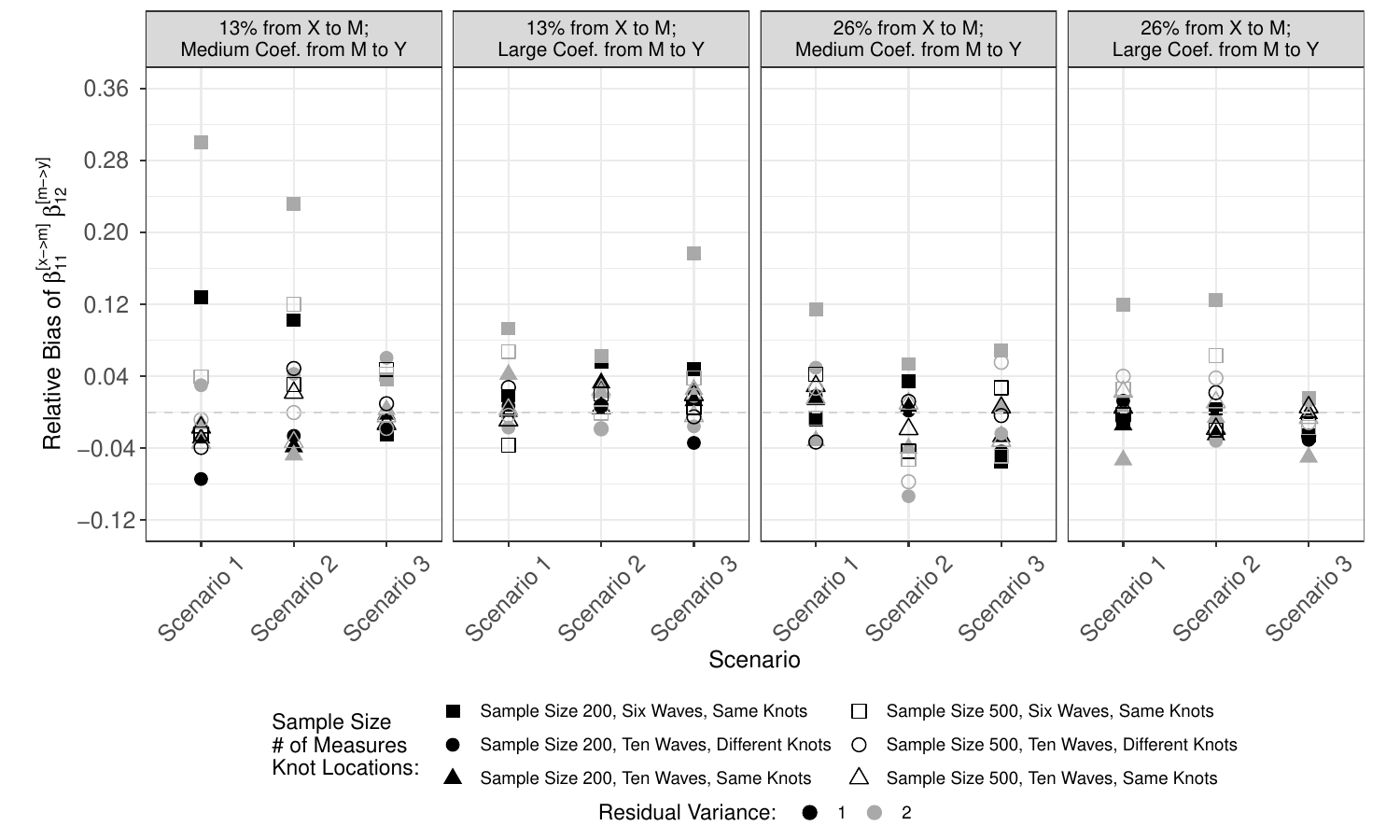}
 \caption{Relative Biases of $x_{1}\rightarrow{m_{1}\rightarrow{y_{2}}}$}
 \label{fig:rBias_x1m1y2}
\end{subfigure}
\begin{subfigure}{0.5\textwidth}
 \centering
 \includegraphics[width=\linewidth]{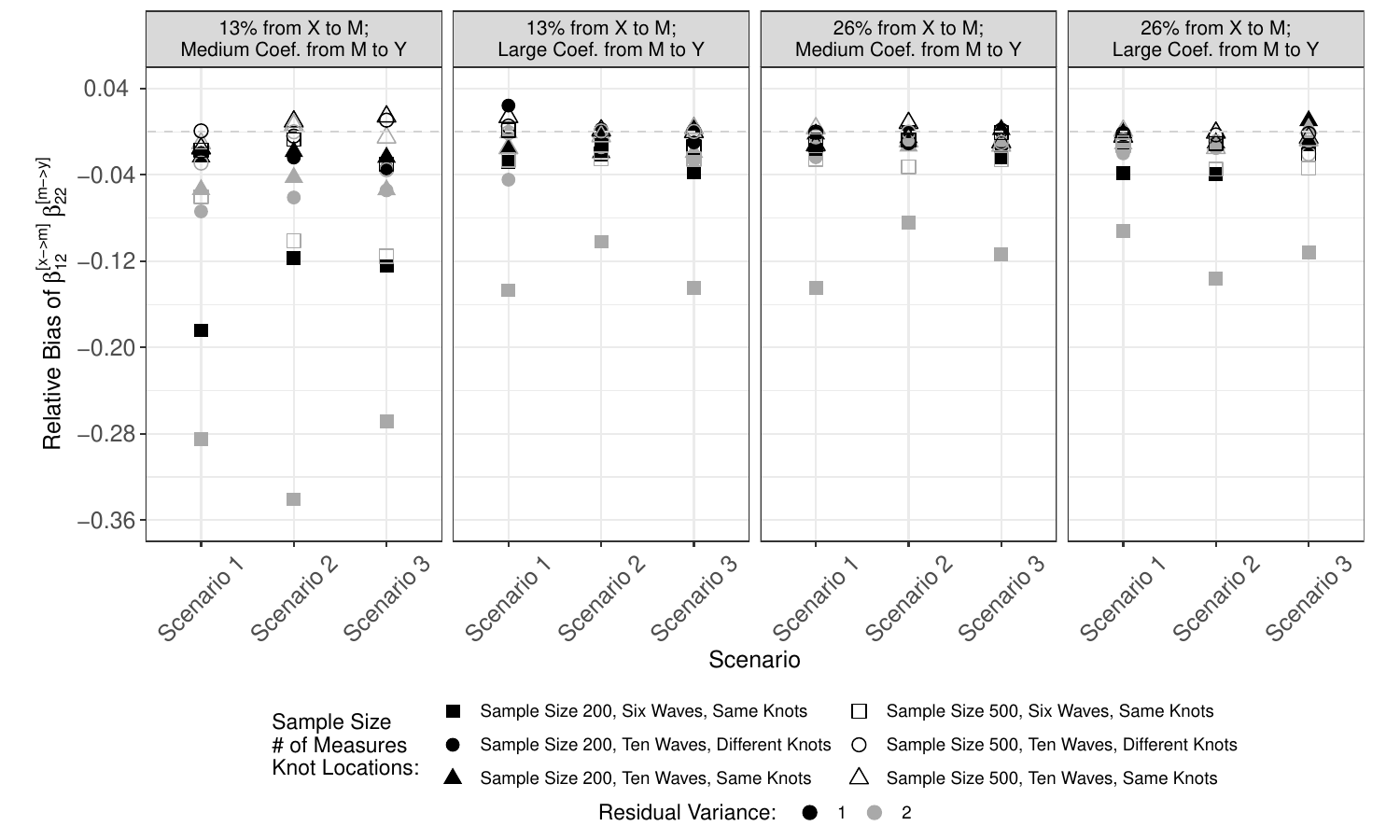}
 \caption{Relative Biases of $x_{1}\rightarrow{m_{2}\rightarrow{y_{2}}}$}
 \label{fig:rBias_x1m2y2}
\end{subfigure}%
\begin{subfigure}{0.5\textwidth}
 \centering
 \includegraphics[width=\linewidth]{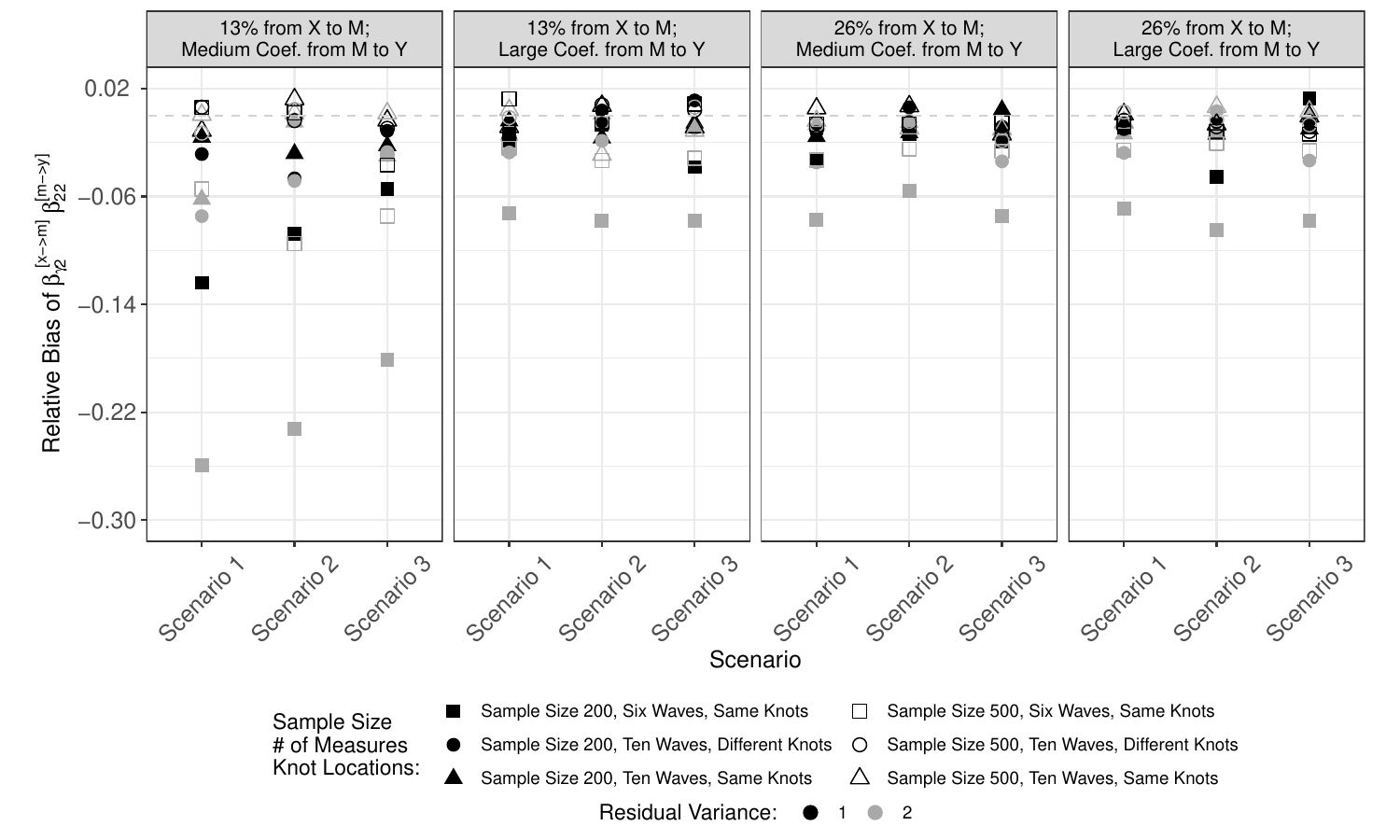}
 \caption{Relative Biases of $x_{\gamma}\rightarrow{m_{2}\rightarrow{y_{2}}}$}
 \label{fig:rBias_xrm2y2}
\end{subfigure}
\begin{subfigure}{0.5\textwidth}
 \centering
 \includegraphics[width=\linewidth]{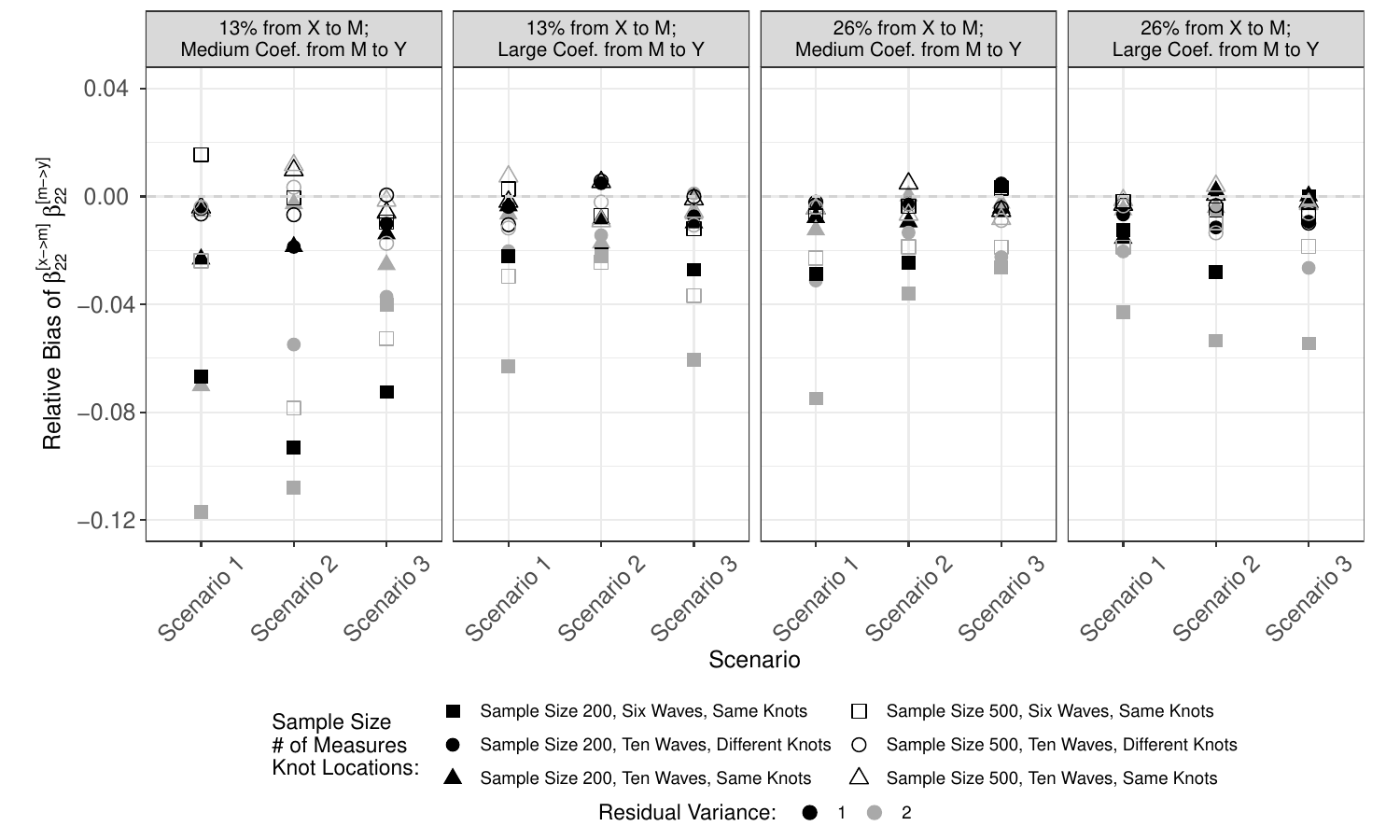}
 \caption{Relative Biases of $x_{2}\rightarrow{m_{2}\rightarrow{y_{2}}}$}
 \label{fig:rBias_x2m2y2}
\end{subfigure}
\caption{Summary of Relative Biases for Mediated Effects with Some Bias Greater than $10\%$ for Model 2}
\label{fig:rBias_M2_med}
\end{figure}

\begin{figure}
\centering
\begin{subfigure}{.5\textwidth}
 \centering
 \includegraphics[width=1.0\linewidth]{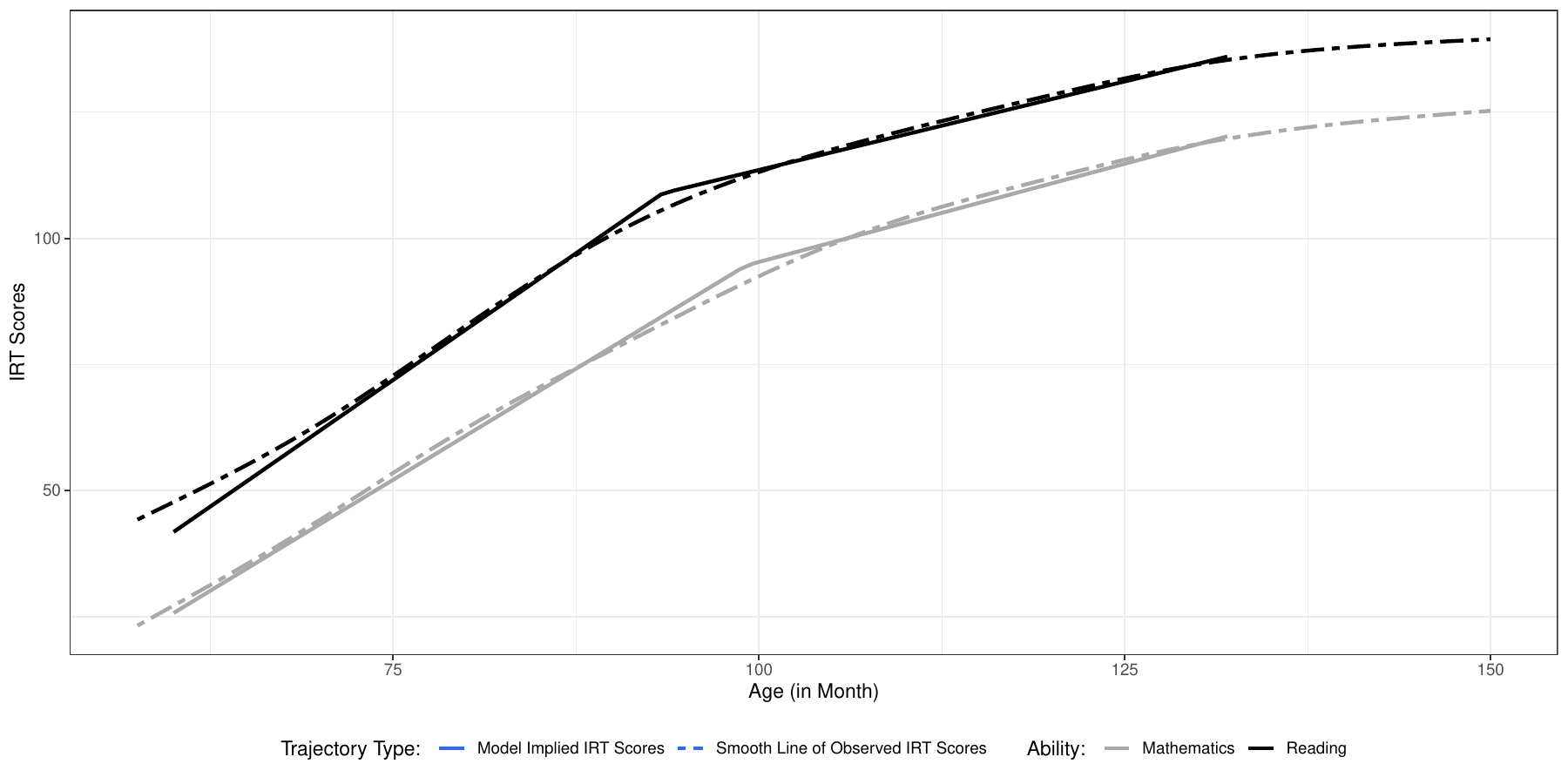}
 \caption{First Longitudinal Mediation Model}
 \label{fig:traj_model1}
\end{subfigure}%
\begin{subfigure}{.5\textwidth}
 \centering
 \includegraphics[width=1.0\linewidth]{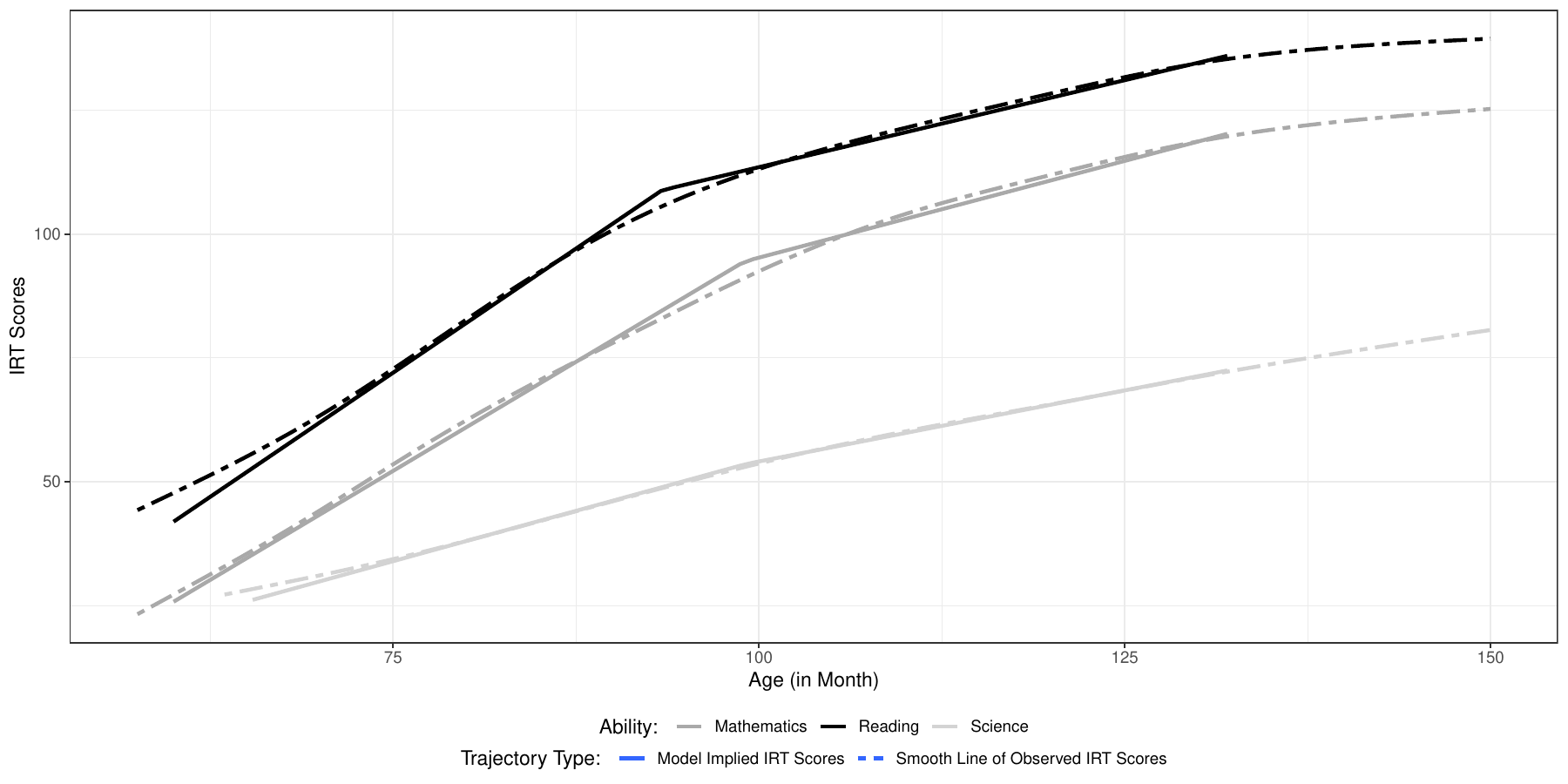}
 \caption{Second Longitudinal Mediation Model}
 \label{fig:traj_model2}
\end{subfigure}
\caption{Model-Implied Trajectory and Smooth Line of Development of Academic Abilities}
\label{fig:traj_models}
\end{figure}
\end{document}